\documentclass[fleqn,usenatbib]{mnras}

\usepackage[T1]{fontenc}

\DeclareRobustCommand{\VAN}[3]{#2}
\let\VANthebibliography\thebibliography
\def\thebibliography{\DeclareRobustCommand{\VAN}[3]{##3}\VANthebibliography}

\usepackage{graphicx}	
\usepackage{amsmath}	
\usepackage{amssymb}	

\usepackage{newtxtext,newtxmath}

\title[IMBH demographics in dwarf galaxies]{Dwarf AGNs from Variability for the Origins of Seeds (DAVOS): Intermediate-mass black hole demographics from optical synoptic surveys}

\author[C. J. Burke et al.]{
Colin J. Burke,$^{1,2}$\thanks{E-mail: colinjb2@illinois.edu (CJB)}
Yue Shen,$^{1,3}$
Xin Liu,$^{1,3}$ 
Priyamvada Natarajan,$^{4,5}$
Neven Caplar,$^{6}$\newauthor
\ Jillian M. Bellovary$^{7,8,9}$
and Z. Franklin Wang$^{1}$
\\
$^{1}$Department of Astronomy, University of Illinois at Urbana-Champaign,
1002 W. Green Street, Urbana, IL 61801, USA\\
$^{2}$Center for AstroPhysical Surveys, National Center for Supercomputing Applications, 
1205 W. Clark Street, Urbana, IL 61801, USA\\
$^{3}$National Center for Supercomputing Applications, 
1205 W. Clark Street, Urbana, IL 61801, USA\\
$^{4}$Department of Astronomy, Yale University, 260 Whitney Avenue, New Haven, CT 06511, USA\\
$^{5}$Black Hole Initiative, Harvard University, 20 Garden Street, Cambridge, MA 02138, USA\\
$^{6}$Department of Astrophysical Sciences, Princeton University, 4 Ivy Lane, 08544, Princeton, NJ, USA\\
$^{7}$Department of Physics, Queensborough Community College, City University of New York, 222-05 56th Ave, Bayside, NY, 11364, USA\\
$^{8}$Department of Astrophysics, American Museum of Natural History, Central Park West at 79th Street, New York, NY 10024, USA\\
$^{9}$Department of Physics, Graduate Center, City University of New York, New York, NY 10016, USA
}

\date{Accepted XXX. Received YYY; in original form ZZZ}

\pubyear{2022}

\begin{document}
\label{firstpage}
\pagerange{\pageref{firstpage}--\pageref{lastpage}}
\maketitle

\begin{abstract}
We present a phenomenological forward Monte Carlo model for forecasting the population of active galactic nuclei (AGNs) in dwarf galaxies observable via their optical variability. Our model accounts for expected changes in the spectral energy distribution (SED) of AGNs in the intermediate-mass black hole (IMBH) mass range and uses observational constraints on optical variability as a function of black hole (BH) mass to generate mock light curves. Adopting several different models for the BH occupation function, including one for off-nuclear IMBHs, we quantify differences in the predicted local AGN mass and luminosity functions in dwarf galaxies. As a result, we are able to model the fraction of variable AGNs as a function of important galaxy host properties, such as host galaxy stellar mass, in the presence of selection effects. We find that our adopted occupation fractions for the ``heavy'' and ``light'' initial BH seeding scenarios can be distinguished with variability at the $2-3 \sigma$ level for galaxy host stellar masses below $\sim 10^8 M_\odot$ with data from the upcoming Vera C. Rubin Observatory. We also demonstrate the prevalence of a selection bias whereby recovered IMBH masses fall, on average, above the predicted value from the local host galaxy - BH mass scaling relation with the strength of this bias dependent on the survey sensitivity. Our methodology can be used more broadly to calibrate AGN demographic studies in synoptic surveys. Finally, we show that a targeted $\sim$ hourly cadence program over a few nights with the Rubin Observatory can provide strong constraints on IMBH masses given their expected rapid variability timescales.
\end{abstract}





\section{Introduction} \label{sec:intro}

Understanding the population of active galactic nuclei (AGN) in the local Universe can provide insights into the growth and evolution of supermassive black holes (SMBHs) across cosmic time. While virtually every massive galaxy contains a SMBH in its center, the occupation fraction of black holes in the dwarf galaxy regime remains poorly constrained. There is however growing strong evidence for the existence of $\sim 10^5 - 10^6M_{\odot}$ black holes in dwarf galaxies \citep{Filippenko2003,Barth2004,Reines2015,Baldassare2015}. However, with the exception of the recent gravitational wave event GW190521 with a merger remnant mass of $142^{+28}_{-16}\ M_{\odot}$ \citep{LIGOVirgo2020}; the X-ray tidal disruption event 3XMM J215022.4-055108 \citep{Lin2018}; and the somewhat more controversial $M_{\rm{BH}} \sim 10^4\ M_{\odot}$ hyper-luminous X-ray source ESO 243-49 HLX-1 \citep{Farrell2009}, intermediate-mass black holes (IMBHs) with $M_{\rm{BH}} \sim 10^2 - 10^4M_{\odot}$ remain difficult to identify \citep{Greene2020}.

To explain these observations, as well as the formation of SMBHs at high redshifts when the Universe was only a few hundred Myr old (e.g., \citealt{Fan2001,Wu2015,Banados2018,Wang2021}), it is thought that SMBHs must grow via accretion and mergers from early seed black holes (e.g., \citealt{Natarajan2014,Inayoshi2020}). Theories of SMBH seeding scenarios broadly fall into two classes: ``light'' and ``heavy'' seeds. In the most popular light seed scenario, black holes with masses of $\sim 10^{1-2}\ M_{\odot}$ are expected to form as remnants of the massive, first generation of stars, namely the Population III (Pop III) stars  \citep{Bond1984,Madau2001,Fryer2001,Abel2002,Bromm2003}. With improvement in the resolution of simulations that track the formation of first stars, it is now found that rather than forming individual stars, early star formation results in star clusters, whose evolution could also provide sites for the formation of light initial seeds \citep{Gurkan2004,PortegiesZwart2004}. Essentially, the light seed scenarios refer to starting with low mass seeds and gradually growing them over time. On the other hand, in the most popular ``heavy'' seed scenario, black holes with masses of $\sim 10^4 - 10^6\ M_{\odot}$ are expected to viably form from direct collapse of primordial gas clouds under specific conditions \citep{Haehnelt1993,Loeb1994,Bromm2003,Koushiappas2004,Lodato2006,Begelman2006,LodatoPN2008} accompanied by accelerated early growth at super-Eddington accretion rates. Additionally multiple other formation channels have also been proposed, such as mechanisms within nuclear star clusters \citep{Devecchi2009,Davies2011,Devecchi2010,Tal2014,Lupi2014,Antonini2015,Stone2017,Fragione2020,Kroupa2020,Natarajan2021}; inside globular clusters \citep{Miller2002,Leigh2014,Antonini2019}; and even young star clusters \citep{Rizzuto2021}. Heavy seeds are predicted to be fewer in number, while light seeds are predicted to be more abundant but less massive \citep{LodatoPN2008}. Given that the host galaxy stellar mass appears to be correlated to the mass of both inactive and active central black holes (BHs) at least in the local Universe; \citealt{Magorrian1998,Reines2015}), the occupation fraction (i.e., fraction of galaxies containing a central BH at a given stellar mass) is expected be a potential observational tracer of seeding \citep{Volonteri2008,Greene2012}. Counter-intuitively, despite their complex growth history via accretion and mergers, the local occupation fraction in the dwarf galaxy mass range ($M_{\star} \lesssim 10^{9.5}\ M_{\odot}$) is predicted to be particularly sensitive to early seeding physics (but see \citet{Mezcua2019Nat}). Even at late cosmic times and on these small dwarf galaxy scales, estimates of the occupation fraction might permit discriminating between the light and heavy seeding scenarios \citep{Volonteri2008,Ricarte2018}.

Deep X-ray surveys have been successfully used to identify low-mass and low-luminosity AGNs at low and intermediate redshifts \citep{Fiore2012,Young2012,Civano2012,Miller2015,Mezcua2016,Luo2017,Xue2017}. However, these surveys are expensive time-wise and are often plagued by contamination from X-ray binaries. Radio searches have also successfully identified low-mass AGNs as radio cores in star-forming dwarf galaxies \citep{Mezcua2019,Reines2020}, although they are subject to low detection rates. Traditional AGN search techniques at optical wavelengths, such as narrow-emission line diagnostics \citep{Baldwin1981,Veilleux1987}, on the other hand tend to miss a large fraction of IMBHs preferentially in star-forming \citep{Baldassare2016,Trump2015,Agostino2019} and low-metallicity \citep{Groves2006} host galaxies. However, systematic searches using wide-area optical surveys have begun to uncover this previously-hidden population of accreting BHs in dwarf galaxies. One popular technique that has been pursued is the mining of large databases of optical spectra for broad emission features in Balmer emission lines \citep{Greene2007,Chilingarian2018,Liu2018}. However, this method requires high $S/N$ spectra to detect the very low-luminosity broad emission \citep{Burke2021c}. In addition, it suffers from contamination from supernovae and stellar winds, which can both produce transient broad Balmer emission with luminosities identical to a dwarf AGN. Confirmation of the detection of dwarf AGN further requires multi-epoch spectroscopy to ensure the broad emission is persistent \citep{Baldassare2016}. Finally, it has been suggested that some accreting IMBHs may fail to produce a broad line region at all \citep{Chakravorty2014}.

The possibility that some IMBHs live outside their host galaxy nuclei---the so called ``wandering'' BH population---is another complicating factor for systematic searches of IMBHs that have traditionally focused on detecting central sources \citep{Volonteri2005,Bellovary2010,Mezcua2015,Mezcua2020,Bellovary2019,Reines2020,Ricarte2021a,Ricarte2021b,Ma2021}. As recently demonstrated from the analysis of the Romulus suite of simulations \citep{Ricarte2021a} demonstrate that a variety of dynamical mechanisms could result in a population of wandering IMBHs in galaxies, such as tidal stripping of merging dwarf galaxies \citep{Zinnecker1988}; gravitational recoil from galaxy centers \citep{Volonteri2003,Holley-Bockelmann2008,OLeary2009,Blecha2011,Blecha2016}, and gravitational runaway processes in star clusters \citep{Miller2002,PortegiesZwart2002,Fragione2018}.

Recently, searches for optical variability in wide-area optical surveys have uncovered hundreds of dwarf AGN candidates \citep{Baldassare2018,Baldassare2020,Burke2021b,Martinez-Palomera2020,Ward2021}. These sources have enabled studies that have improved our understanding of AGN optical variability across a vast range of mass scales. Variability is thought to be driven by the inner UV-emitting regions of their rapidly-accreting accretion disks \citep{Burke2021}. In this work, we leverage these recent advances in IMBH identification and optical variability behavior, along with extrapolations of known host-galaxy correlations observed in the low-mass regime (e.g., \citealt{Reines2015}), to forecast the IMBH population that could be detectable by upcoming time-domain imaging surveys.

Our paper is organized as follows. In \S\ref{sec:model}, we develop a forward model to forecast the number density of IMBHs in dwarf galaxies. In \S\ref{sec:obs}, we adapt this model to generate simulated observations mimicking light curves expected from the Vera C. Rubin Observatory Legacy Survey of Space and Time (LSST Rubin; \citealt{Ivezic2019}) and the Palomar Transient Factory (PTF) survey \citep{Law2009} to compare with existing observations \citep{Baldassare2020}. We opt for the PTF comparison over a similar study with SDSS \citep{Baldassare2018} because the PTF study has a larger sample size which enables tighter constraints on the variable fraction while being broadly consistent with the SDSS data. A comparison with the Dark Energy Survey is presented separately in \citet{Burke2021b}. We demonstrate the capability of our model to reproduce the IMBH detection fraction as a function of stellar mass consistent with existing AGN demographic studies. A concordance $\Lambda$CDM cosmology with $\Omega_{m} = 0.3$, $\Omega_{\Lambda} = 0.7$, and $H_0 = 70$ km s$^{-1}$ Mpc$^{-1}$ is assumed throughout. Unless stated otherwise, all uncertainty bands in the figures are $1\sigma$, estimated using the 16th and 84th percentiles of the probability density distributions, and points are the distribution means. Duplicate symbols are used for some parameters throughout this work. The reader is requested to refer to the context to resolve any ambiguity.

\section{Methodology to construct the demographic model}\label{sec:model}

\begin{figure*}
\includegraphics[width=1\textwidth]{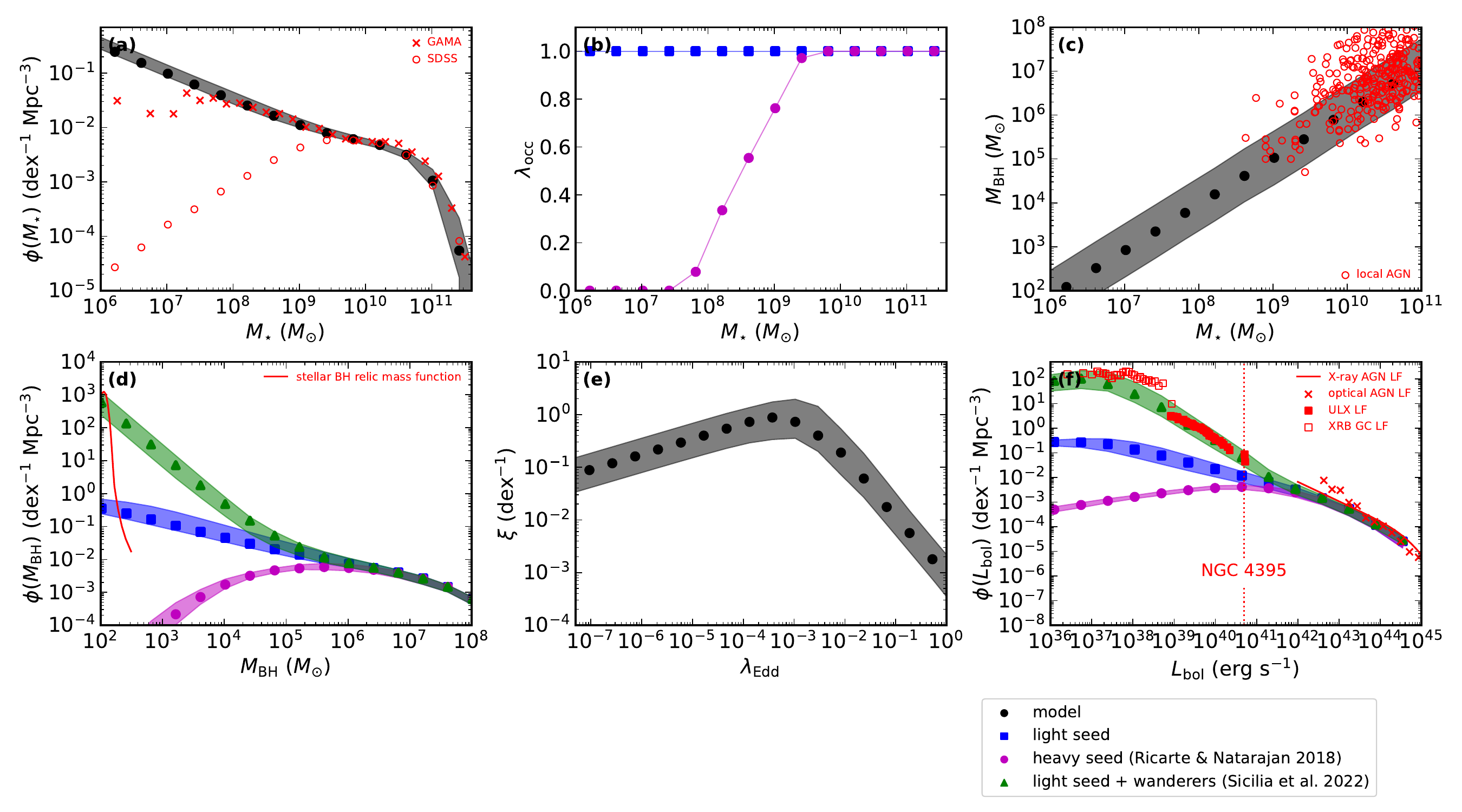}
\caption{Monte Carlo model for local AGN demographics in dwarf galaxies. We start from the galaxy stellar mass function (a) and consider two possibilities for the occupation fraction (a ``light'' seed scenario in blue/square symbols and a ``heavy'' seed scenario in magenta/circle symbols; b). Then, we use the local $M_{\rm{BH}}-M_{\star}$ scaling relation (c) to predict the BH mass function (d). Finally, we assume a power-law distribution for the Eddington ratios (e) to predict the local bolometric AGN luminosity function (LF) (f). The shaded bands are $1\sigma$ uncertainties, estimated using the 16th and 84th percentiles of the distributions, and points are the distribution means. The red `x' symbols are the observational constraints on the local galaxy stellar mass function (points below $\sim10^8\ M_{\odot}$ are effected by incompleteness of low surface brightness galaxies; \citealt{Baldry2012}) (a) and the local AGN luminosity function from optical observations \citep{Schulze2009,Hao2005} (f). The red circles in panel (c) are the sample of local broad-line AGNs from \citet{Reines2015}. The red curve in panel (d) is stellar BH relic mass function anchored to merger rates from gravitational wave observations \citep{Sicilia2022}. The red line in panel (f) is the best-fit broken power law to the local AGN luminosity function derived from X-ray observations \citep{Ajello2012}. The red `o' symbols is the GSMF measured from the SDSS-based NASA Sloan Atlas catalog \citep{Blanton2011}, which demonstrates the spectroscopic incompleteness at low stellar mass (a). The red filled squares in panel (f) is the observed luminosity function of ultra-luminous X-ray sources derived from seven collisional ring galaxies \citep{Wolter2018} normalized to the number density of $M_\star \sim 10^6\ M_{\odot}$ dwarf galaxies after excluding sources with $L_{0.5-10\ {\rm{keV}}} < 10^{39}$ erg s$^{-1}$ where the sample is incomplete. The red open squares in panel (f) is the observed luminosity function of X-ray binaries (XRBs) in globular clusters (GCs) in nearby galaxies \citep{Lehmer2020} normalized to the number density of $M_\star \sim 10^6\ M_{\odot}$ dwarf galaxies. The red dotted vertical line in panel (f) represents the bolometric luminosity of the $M_{\rm{BH}} \sim 10^4-10^5\ M_{\odot}$ dwarf Seyfert galaxy NGC 4395 \citep{Filippenko2003,Moran1999}. \label{fig:model}}
\end{figure*}


Broadly following the basic methodology presented in prior work by \citet{Caplar2015} and \citet{Weigel2017}, we develop an empirically motivated forward model starting from the galaxy stellar mass function and host-galaxy scaling relations to derive the corresponding BH mass and AGN luminosity functions (also see \citealt{Gallo2019,Greene2020}). Our goal is to estimate the number density of dwarfs with central AGNs in the IMBH mass range that would result from the various proposed seeding mechanisms. Therefore, we must extrapolate scaling relations derived from current observational constraints on the galaxy population from host galaxy correlations as well as the Eddington ratio distribution derived for more massive AGNs to lower mass BHs. A summary table of parameters and our adopted values for them are provided in Table~\ref{tab:par}, unless otherwise explicitly quoted in the text.

\subsection{The dwarf galaxy population}
\label{sec:GSMF}

\begin{table}
\centering
\caption{Table of parameters, our adopted values and their $1\sigma$ uncertainties describing the galaxy population of our Monte Carlo model.}
\label{tab:par}
\small
\begin{tabular}{cccc}
\hline \hline
Parameter & Value & Unit & Reference \\
\hline
\multicolumn{4}{|c|}{Galaxy Stellar Mass Function (GSMF)}\\
\hline
$\log(M_{\star}^{\ast}/M_{\odot})$ & $10.78 \pm 0.01$ & dex & \citet{Wright2017} \\
$\phi_1/10^{-3}$ & $2.93 \pm 0.40$ & Mpc$^{-3}$ & \ldots \\
$\phi_2/10^{-3}$ & $0.63 \pm 0.10$ & Mpc$^{-3}$ & \ldots \\
$\alpha_1$ & $-0.62 \pm 0.03$ &  & \ldots \\
$\alpha_2$ & $-1.50 \pm 0.01$ &  & \ldots \\
\hline
\multicolumn{4}{|c|}{$^{a,b}$Blue$+$Green Galaxy Stellar Mass Function (GSMF)}\\
\hline
$\log(M_{\star}^{\ast}/M_{\odot})$ & $10.72$ & Mpc$^{-3}$ & \citet{Baldry2012} \\
$\phi/10^{-3}$ & $0.71$ &  & \ldots \\
$\alpha$ & $-1.45$ &  & \ldots \\
\hline
\multicolumn{4}{|c|}{$^{a}$Red Galaxy Stellar Mass Function (GSMF)}\\
\hline
$\log(M_{\star}^{\ast}/M_{\odot})$ & $10.72$ & dex & \citet{Baldry2012} \\
$\phi_1/10^{-3}$ & $3.25$ & Mpc$^{-3}$ & \ldots \\
$\phi_2/10^{-3}$ & $0.08$ & Mpc$^{-3}$ & \ldots \\
$\alpha_1$ & $-0.45$ &  & \ldots \\
$\alpha_2$ & $-1.45$ &  & \ldots \\
\hline
\multicolumn{4}{|c|}{$^{c}$Host Galaxy-Black Hole Mass Scaling}\\
\hline
$\log(M_{\star}^{\ast}/M_{\odot})$ & $11$ & dex & \citet{Reines2015} \\
$\alpha$ & $7.45 \pm 0.08$ &  &  \ldots \\
$\beta$ & $1.05 \pm 0.11$ &  & \ldots \\
\hline
\multicolumn{4}{|c|}{Blue$+$Green Eddington Ratio Distribution Function (ERDF)}\\
\hline
$\log(\lambda^{\ast}_{\rm{Edd}})$ & $-1.84^{+0.30}_{-0.37}$ &  & \citet{Weigel2017} \\
$\delta_1$ & $-0.2$ &  & $^{d}$ \\
$\delta_2$ & $2.53^{+0.68}_{-0.38}$ &  & \citet{Weigel2017} \\
$\log(\lambda_{\rm{Edd, min}})$ & $-8$ &  & \\
$\log(\lambda_{\rm{Edd, max}})$ & $0$ &  & \\
\hline
\multicolumn{4}{|c|}{Red Eddington Ratio Distribution Function (ERDF)}\\
\hline
$\log(\lambda^{\ast}_{\rm{Edd}})$ & $-2.84^{+0.22}_{-0.14}$ &  & \citet{Weigel2017} \\
$\delta_1$ & $-0.3$ &  & $^{d}$ \\
$\delta_2$ & $1.22^{+0.19}_{-0.13}$ &  & \citet{Weigel2017} \\
$\log(\lambda_{\rm{Edd, min}})$ & $-8$ &  &  \\
$\log(\lambda_{\rm{Edd, max}})$ & $0$ &  & \\
\hline
\end{tabular}\\
{\raggedright
$^{a}$ We use the \citet{Wright2017} GSMF, which is better-constrained in the dwarf galaxy regime, but use the separate blue$+$green and red GSMFs from \citet{Baldry2012} to determine the relative ratio of the blue$+$green and red galaxy populations (see text for details). \\
$^{b}$ This is a single \citet{Schechter1976} function in \citet{Baldry2012}. \\
$^{c}$ We adopt the rms scatter in the relation of $\sim 0.55$ dex in the $M_{\star}$ direction \citep{Reines2015}. \\
$^{d}$ We re-normalized the $\delta_1$ parameters to better approximate the variable fraction of the entire galaxy population. Our normalization is still consistent with the local AGN luminosity function. \\
\par}
\end{table}

We begin by considering the number density of galaxies in the local Universe. At a given redshift, the measured galaxy stellar mass function (GSMF) is well-described by a double power-law function of the form,
\begin{equation}
\label{eq:GSMF}
    \phi(M_{\star})\ dM_{\star} = e^{-M_{\star}/M_{\star}^{\ast}}\ \left[\phi_1\left(\frac{M_{\star}}{M_{\star}^{\ast}}\right)^{\alpha_1} + \phi_2\left(\frac{M_{\star}}{M_{\star}^{\ast}}\right)^{\alpha_2}\right] \frac{dM_{\star}}{M_{\star}^{\ast}},
\end{equation}
where $\phi=dn_{\star}/dM_{\star}$, $M_{\star}$ is the galaxy stellar mass, $n_{\star}$ is the number density, $M_{\star}^{\ast}$ is the break stellar mass, $\alpha_1$ and $\alpha_2$ are the shallow and steep power law exponents, respectively, and $\phi_1$ and $\phi_2$ are normalization factors that correspond to the low and high mass end of the GSMF, respectively \citep{Schechter1976} . We adopt the best-fit parameters from \citet{Wright2017} based on the Galaxy And Mass Assembly (GAMA) low-redshift $\sim$180 deg$^2$ spectroscopic survey, which has a spectroscopic depth of $r\sim19.8$ mag \citep{Driver2011,Liske2015}. The GAMA survey measured GSMF is good to $z\sim0.1$ and for $M_{\star} \gtrsim 10^{7.5}\ M_{\odot}$ but is also consistent with current limits on the GSMF down to $M_{\star} \sim 10^{6.5}\ M_{\odot}$ from deep G10-COSMOS imaging---a $\sim$1 deg$^2$ subset of the GAMA survey overlapping with the Cosmic Evolution Survey \citep{Scoville2007} with a spectroscopic depth of $r\sim24.5$ mag \citep{Andrews2017}.

The high mass end of the GSMF is mostly constituted by red galaxies, while the low mass end of the GSMF is dominated by blue galaxies. Although the \citet{Wright2017} parameters are well-constrained for the low-mass end of the GSMF, they do not include separate derived GSMFs and tailored fits for the red and blue galaxy populations. Therefore, we use the ratio of the GSMFs partitioned between the red and blue galaxy populations from \citet{Baldry2012}, which is consistent with the results of \citet{Wright2017}, to separately populate red and blue galaxies in our model. We assign each galaxy a ``red'' or ``blue'' identifier, which we use to determine the accretion mode, that differs between these two galaxy populations \citep{Weigel2017,Ananna2022}. We ignore any redshift dependence in the GSMF, as we show that the number of detectable IMBHs drops off quickly with redshift at the expected sensitivities ($g\sim25$ mag) for LSST Rubin currently being considered. Our LSST model-predicted, detectable IMBHs are expected to mostly lie at $z\lesssim 0.05$.

The number of random draws $N_{\rm{draw}}$ can be defined in terms of the GSMF and the survey volume as:
\begin{equation}
\label{eq:Ndraw}
    N_{\rm{draw}} = V(z_{\rm{min}}, z_{\rm{max}}, \Omega)\ \int_{M_{\star,\rm{min}}}^{ M_{\star,\rm{max}}} \phi({M_\star})\ dM_\star,
\end{equation}
where $V$ is comoving volume between redshifts $z_{\rm{min}}$ and $z_{\rm{max}}$ over solid angle $\Omega$. With each draw we randomly assign $N_{\rm{draw}}$ galaxies a stellar mass using Equation~\ref{eq:GSMF} as the target distribution with $z_{\rm{min}}=0$, $z_{\rm{max}}=0.055$. Our choice of $z_{\rm{max}}=0.055$ is chosen to match existing observational constraints \citep{Baldassare2020}, and we show that the number of detectable IMBHs falls off dramatically with increasing redshift. This assumption of the restriction of the redshift range under scrutiny also allows us to ignore any explicit redshift dependence in the GSMF. The galaxy redshifts are determined by randomly assigning each galaxy to a redshift bin out to $z_{\rm{max}}$, where the number of galaxies in each redshift bin is then proportional to the cosmological differential comoving volume at that redshift bin. As a consistency check, we show that the redshift and stellar mass distributions of our mock sample compare extremely well to observed SDSS galaxies in Appendix~\ref{sec:nsacomp}.

\subsection{Occupation fraction}

After determining $N_{\rm{draw}}$, we then consider different possible functional forms for the occupation fraction, the fraction of galaxies hosting an IMBH/SMBH, $\lambda_{\rm{occ}}(M_{\star})$. We refer to this quantity as the \emph{occupation function}. This quantity may be greater than unity if multiple IMBHs are harbored in a galaxy. We explore the following scenarios for the occupation function:
\begin{enumerate}
  \item \textbf{Light seeds:} A constant occupation function of $\lambda_{\rm{occ}} = 1$ shown in blue in Figure~\ref{fig:model}. This represents the most optimistic predictions for an initial ``light'' seeding scenario (e.g., from Pop. III stellar remnants) as examined in \citet{Ricarte2018}.
  \item \textbf{Heavy seeds:} An occupation function that approaches unity for massive galaxies ($M_{\star} > 10^9\ M_{\odot}$) but drops dramatically by $M_{\star} \sim 10^8\ M_{\odot}$, shown in magenta according to the ``heavy-MS'' scenario (e.g., from direct collapse channels) adopted from \citet{Ricarte2018}. This prediction is derived from a semi-analytic model which traces the evolution of heavy seeds under the assumption of a steady-state accretion model that reproduces the observed AGN main-sequence. This resulting occupation fraction is broadly consistent with studies from cosmological simulations \citep{Bellovary2019}.
  \item \textbf{Light seed $+$ wanderers:} We adopt an occupation fraction anchored to the \citet{Sicilia2022} BH mass function (BHMF) derived from ongoing stellar formation channels. The \citet{Sicilia2022} BHMF describes the local IMBH population by anchoring to merger rates derived from gravitational wave (GW) observations by LIGO/VIRGO \citep{Abbott2021,Baxter2021}. We assume a smooth transition between these GW anchors to the \citet{Sicilia2022} BHMF at $M_{\rm{BH}} \sim 10^2\ M_{\odot}$ and the BHMF from scenario (i) at $M_{\rm{BH}} \sim 10^4\ M_{\odot}$ as a reasonable model to approximate the wandering and off-nuclear IMBHs that have not yet fallen to the center of the host galaxy. The resulting occupation fraction is broadly consistent with the existing constraints on the luminosity function derived from AGNs, ultra-luminous X-ray sources (ULXs), and XRBs as shown in Figure~\ref{fig:model}(f).
\end{enumerate}
Scenarios (i) and (ii) both assume a single seeding epoch and subsequent growth of the seed BH to fall onto the black hole-host galaxy mass relation at late times. However, stellar cluster seed formation channels can continuously produce IMBHs as recently pointed by \citet{Natarajan2021}. There are considerable theoretical uncertainties in these models arising from the hitherto unknown efficiencies of continual seed formation processes. We will incorporate continual BH formation models in future work. Furthermore, multiple seeding scenarios could simultaneously be at work in the Universe, and this implies that theoretical constraints on the occupation functions do remain uncertain for this reason as well. In this work, however, we pursue a brand new avenue and explore if optical variability can be used to constrain the occupation function. Precisely how the low-redshift occupation fraction traces seeding scenarios at high redshifts is more complex question that requires more detailed interpretation due to the interplay with accretion physics \citep{Mezcua2019Nat}. Here, we adopt these different scenarios described above as a way to bracket the possible reasonable outcomes.

We are unaware of quantitative predictions for how the occupation fraction or number density of the wandering BH population is connected to the host galaxy stellar mass. Hence, we will not consider scenario (iii) in our investigation of the variable AGN population versus host galaxy properties. We emphasize the need for future study of such theoretical developments in the future when observations start to confirm the existence of a wandering population. For scenario (iii), the number of black holes is simply determined by the BHMF, but we remain agnostic about how which galaxies they live within.

For each of the scenarios (i) and (ii), we assign each galaxy a BH or not according to its occupation probability. Therefore, the remaining number of draws is given by,
\begin{equation}
    N_{\rm{draw, BH}} = N_{\rm{draw}} \int^{M_{\rm{\star,max}}}_{M_{\rm{\star,min}}} \lambda_{\rm{occ}}(M_{\star})\ dM_{\star},
\end{equation}
where $N_{\rm{draw}}$ is given by Equation~\ref{eq:Ndraw}. 


\subsection{Black hole mass scaling relations}

In the local universe, the stellar mass of the AGN host galaxy scales with the mass of the central BH as a power-law of the form:
\begin{equation}
\label{eq:MMstar}
    \log\left(\frac{M_{\rm{BH}}}{M_{\odot}}\right) = \alpha + \beta \log\left(\frac{M_{\star}}{M_{\star}^{\ast}}\right).
\end{equation}

We adopt the relation measured from local broad-line AGNs including dwarf galaxies with $\alpha=7.45\pm0.08$; $\beta=1.05\pm0.11$; with a pivot mass $M_{\star}^{\ast} = 10^{11}\ M_{\odot}$ \citep{Reines2015} to obtain BH masses for scenario (i) and (ii). We also include the rms scatter of $\sim 0.6$ dex in $M_{\rm{BH}}$ in the relation when assigning each galaxy a BH mass.

For the wandering BH population of scenario (iii), we assume an analogous relation between the BH mass and mass of the star cluster containing the IMBH $M_{\rm{SC}}$ to obtain their associated stellar masses:
\begin{equation}
\label{eq:MMNSC}
    \log\left(\frac{M_{\rm{SC}}}{M_{\odot}}\right) = \alpha + \beta \log\left(\frac{M_{\rm{BH}}}{M_{\rm{BH}}^\ast}\right).
\end{equation}
We adopt the best-fit parameters from the relation between the BH mass and mass of the nuclear star cluster (as a proxy for $M_{\rm{SC}}$) derived from low-mass nuclear star clusters by \citet{Graham2020} with $\alpha=7.70\pm0.20$; $\beta=0.38\pm0.06$; $M_{\rm{BH}}^{\ast} = 10^{7.89}\ M_{\odot}$ and an intrinsic scatter of $\sim 0.5$ dex in $M_{\rm{SC}}$. Although by definition wandering black holes would not all necessarily be found in nuclear star clusters, to first order, we assume that this relation offers a reasonable description for off-nuclear star clusters with wandering IMBHs. For the wandering BH population, we will use $M_{\rm{SC}}$ in place of host galaxy stellar mass $M_{\star}$ to compute the luminosity from starlight that dilutes the variability.

\subsection{The Eddington ratio distribution}

We adopt a broken power-law distribution for the Eddington luminosity ratio ($\lambda_{\rm{Edd}} \equiv L_{\rm{bol}}/L_{\rm{Edd}}$) probability distribution function to compute the AGN bolometric luminosity $L_{\rm{bol}}$ from this. Specifically, we adopt the commonly used double power-law parameterization \citep{Caplar2015,Sartori2015,Sartori2019,Weigel2017,Pesce2021,Ananna2022}:
\begin{equation}
\label{eq:ERDF}
    \xi(\lambda_{\rm{Edd}}) = \xi^\ast \left[\left(\frac{\lambda_{\rm{Edd}}}{\lambda_{\rm{Edd}}^\ast}\right)^{\delta_1} + \left(\frac{\lambda_{\rm{Edd}}}{\lambda_{\rm{Edd}}^\ast}\right)^{\delta_2}\right]^{-1},
\end{equation}
where $\xi(\lambda_{\rm{Edd}})$ is the Eddington ratio distribution function (ERDF); $\lambda_{\rm{Edd}}^\ast$ is the break Eddington ratio; and $\delta_1$; $\delta_2$ are the shallow and steep power law exponents, respectively.

There is compelling evidence that the red and blue galaxy populations that host central AGN accrete in different modes. \citet{Weigel2017} found that the radio AGN luminosity function (predominately red host galaxies) are described by a broken power law ERDF favoring lower accretion rates. On the other hand, the X-ray AGN luminosity function (predominately blue host galaxies) described by a broken power law ERDF is found to favor relatively higher accretion rates. \citet{Weigel2017} interpret this as evidence for a mass-independent ERDF for red and blue galaxies with radiatively inefficient and efficient accretion modes, respectively. We adopt the best-fit parameters for the high-end slope and break Eddington ratio for the red and blue galaxy populations $\delta_2$, and $\log\ \lambda^\ast$ from \citet{Weigel2017}, in order to match constraints on the $z\approx0$ AGN bolometric luminosity function (e.g., \citealt{Ajello2012,Aird2015}). In seed scenario (iii), we assume that the wandering IMBH population produced through stellar formation channels anchored to the \citet{Sicilia2022} BHMF are described by the radio AGN ERDF favoring lower accretion rates, which is broadly consistent with expectations that wandering black holes are expected to have lower accretion rates \citep{Bellovary2019,GuoM2020,Ricarte2021a,Seepaul2022}.

The normalization of the ERDF $\xi^\ast$ determines how many of the randomly drawn $N_{\rm{draw, BH}}$ BH mass values are assigned an Eddington ratio. Unlike \citet{Weigel2017}, we wish to consider an ERDF normalization that describes the \emph{entire} red and blue galaxy population (rather than separate classes of radio or X-ray selected AGNs). Therefore, our ERDFs must be re-normalized accordingly. We set $\xi^\ast$ such that the integral of the ERDF from $\lambda_{\rm{Edd, min}}$ to $\lambda_{\rm{Edd, max}}$ is 1. This means that all $N_{\rm{draw, BH}}$ BH values are assigned an Eddington ratio and we have assumed that it is independent of BH mass. Then, noting that the low-end slope $\delta_1$ is not well-constrained by the AGN luminosity function for $\delta_1 < -\alpha_1$ (the low-luminosity end of the luminosity function is then determined by $\alpha_1$; \citealt{Caplar2015}). We allow $\alpha_1$  to be a free parameter in our model and adjust it to match the overall variable AGN fraction while maintaining consistency with the AGN luminosity function.

The best-fit parameters for radiatively-efficient AGNs from \citet{Weigel2017} are consistent with the ERDF for low-mass galaxies from \citet{Bernhard2018}. Radiatively-efficient, low-mass AGNs dominate in number, and have the largest impact on the luminosity function. Although alternative ERDFs have been proposed \citep{Kauffmann2009}, the simple mass-independent broken power-law function is able to adequately reproduce observations once selection effects are accounted for \citep{Jones2016,Ananna2022}. Finally, we caution that a population of $z\sim0$ X-ray obscured Compton thick AGNs may be missing from our entire census and hence absent in the luminosity function as well. We consider the optically-obscured AGN fraction later on in this work before computing the optical-band luminosities.

\subsection{Model consistency with observational constraints}

A schematic detailing our model results using random sampling is shown in Figure~\ref{fig:model}. To ensure that our model parameters are consistent with all available relevant observational constraints, we compare our model AGN luminosity function to the observed local AGN luminosity function from \citet{Hao2005} and \citet{Schulze2009} measured using Type 1, broad-line AGNs from the Sloan Digital Sky Survey (SDSS; faint end) and the Hamburg/ESO Survey (bright end). The number densities in each bin $i$ are given by:
\begin{equation}
    \phi_i(x) = \frac{n_i}{V(z_{\rm{min}}, z_{\rm{max}}, \Omega) \times \Delta \log x},
\end{equation}
where $x$ is substituted for the variable of interest e.g., $M_{\star}$, $M_{\rm{BH}}$, or $L_{\rm{bol}}$. We fix the ERDF parameters to reproduce the observed local AGN luminosity function from \citet{Ajello2012} starting with the best-fit parameters of \citet{Weigel2017} and re-normalizing the ERDF to describe the entire galaxy population. This is in reasonably good agreement with the Type 1 bolometric $z\approx0$ AGN luminosity function \citep{Schulze2009,Hao2005}. We separately consider a Type 1/Type 2 AGN fraction before computing the observable optical luminosities for the AGN population.

To check for the consistency of our derived luminosity functions with observations at luminosities below $\sim10^{42}$ erg s$^{-1}$, we show the observed luminosity function of ULXs derived from \emph{Chandra} observations of seven collisional ring galaxies \citep{Wolter2018}. ULXs are non-nuclear sources with X-ray luminosities in excess of $10^{39}$ erg s$^{-1}$, generally thought to be X-ray binaries or neutron stars accreting at super-Eddington rates. However, it is possible that some ULXs are in fact accreting IMBHs (e.g., as noted in \citealt{Feng2011,Kaaret2017}). Regardless, it is important to check that our model bolometric luminosity function for the wandering IMBH population does not exceed the luminosity functions derived from ULXs as a limiting case. We demonstrate the consistency in Figure~\ref{fig:model}, by assuming a bolometric correction factor of $1.25$ \citep{Anastasopoulou2022}. We exclude sources with X-ray luminosities below $10^{39}$ erg s$^{-1}$, where the sample is incomplete \citep{Wolter2018}. We normalize the \citet{Wolter2018} (per-galaxy) luminosity function to the number density of $M_\star \sim 10^{6}\ M_{\odot}$ ultra-low mass dwarf galaxies of $\sim 10^{-1}$ Mpc$^{-3}$ \citep{Baldry2012}, whose IMBHs should dominate the low luminosity end of the BH luminosity function. This comparison should be treated with caution, because the \citet{Wolter2018} sample of massive collisional ring galaxies are not fully representative of all dwarf galaxies, and the normalization of the luminosity function is expected to depend on the star formation rate of the host galaxy \citep{Grimm2003}.

In addition to ULXs, we show the completeness-corrected luminosity function of X-ray binaries (XRBs) spatially coincident with globular clusters (GCs) in nearby galaxies from \citet{Lehmer2020} assuming a bolometric correction of $1.25$ \citep{Anastasopoulou2022}. Again, we confirm that our predicted luminosity functions do not significantly exceed the observed luminosity function of XRBs in GCs after normalizing the luminosity function to the number density of $M_\star \sim 10^{6}\ M_{\odot}$ ultra-low mass dwarf galaxies. Similar caveats exist with this comparison and with that of the ULXs, as the results are also expected to depend on the properties of the star cluster.

As an additional check, we plot the GSMF measured from the SDSS-based NASA Sloan Atlas\footnote{\url{http://nsatlas.org/data}} catalog \citep{Blanton2011} of $z<0.055$ galaxies, which serves as the parent sample of the existing observational constraints \citep{Baldassare2018,Baldassare2020} using the spectroscopic survey area of $\Omega\approx9380$ deg$^2$ (see \citealt{Weigel2016}). The SDSS GSMF is roughly consistent with the \citet{Wright2017} GSMF above $M_{\star} \sim 10^{10}\ M_{\odot}$ but is highly incomplete below. Deeper catalogs will be required to take advantage of the next generation of optical time-domain imaging surveys.

\subsection{Optical bolometric corrections}\label{sec:sed}

\begin{figure*}
\centering
\includegraphics[width=0.85\textwidth]{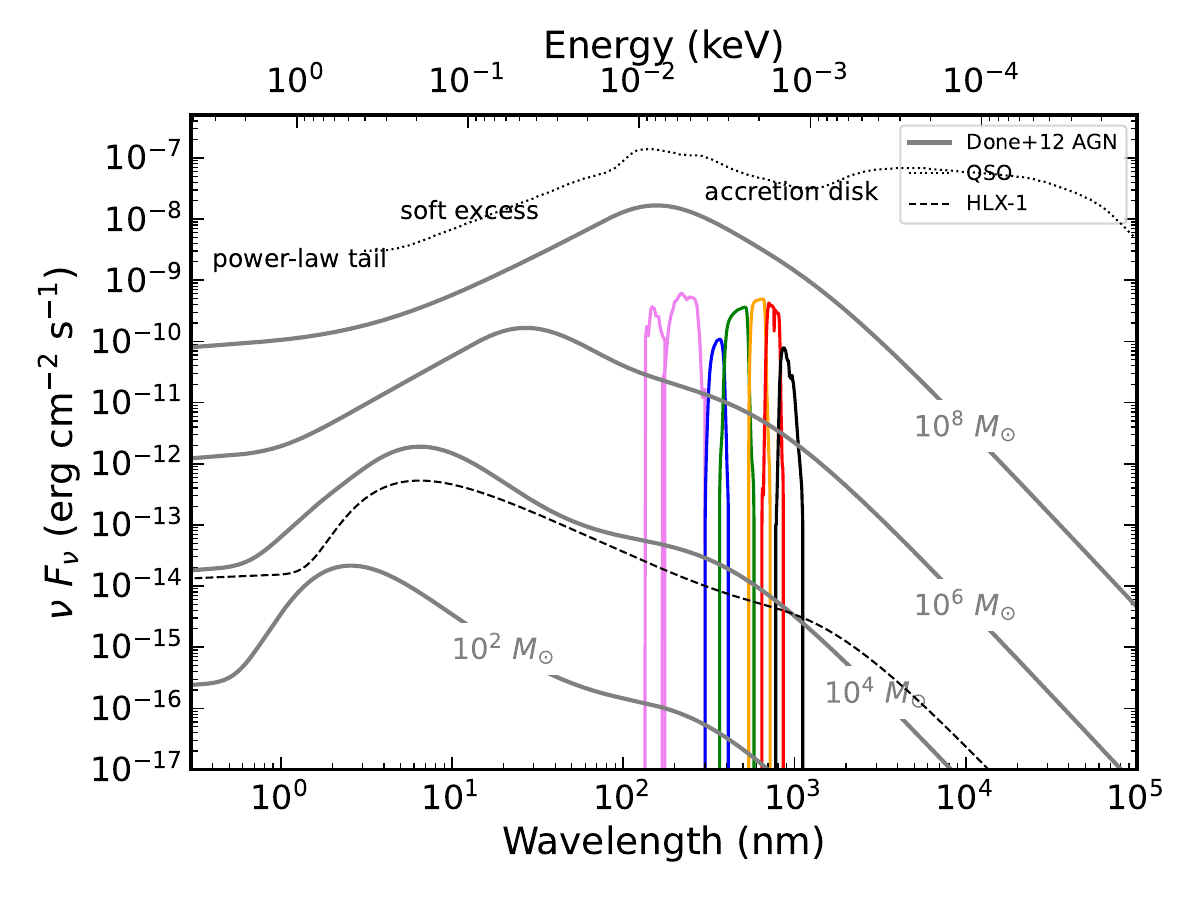}
\caption{Example spectral energy distributions (SEDs) of AGNs with BH masses in the range $M_{\rm{BH}} = 10^2 - 10^8\ M_{\odot}$ with $L_{\rm{bol}}/L_{\rm{Edd}} = 0.1$ using the model of \citet{Done2012} (thick gray lines, denoted ``Done+12 AGN''). We assume a distance of $30$ Mpc for these models. We also show the filter transmittance (throughput) curves for the GALEX (FUV and NUF; violet) and SDSS bandpasses ($ugriz$; blue to black) for reference (arbitrary $y$-axis scaling). We label the approximate locations of the dominant SED component in black text (but note the shift of their peak wavelengths to the left as $M_{\rm{BH}}$ decreases). The dashed black line is the best-fit \citet{Gierlinski2009} irradiated disk model of the IMBH candidate HLX-1 ($M_{\rm{BH}} \sim 10^{4}\ M_{\odot}$; \citealt{Farrell2014}), re-scaled to a distance of 30 Mpc and $L_{\rm{bol}}/L_{\rm{Edd}} \sim 0.1$ for comparison (denoted ``HLX-1''). The dotted black line is the Type 1 quasar SED of \citet{Richards2006} for SMBHs derived from composite observations (denoted ``QSO''). Note the \citet{Richards2006} SED contains emission from the AGN torus at $>1$ microns (i.e., the IR bump), while the \textsc{xspec} SEDs do not contain the torus emission. \label{fig:sed}}
\end{figure*}

\begin{figure*}
\centering
\includegraphics[width=0.85\textwidth]{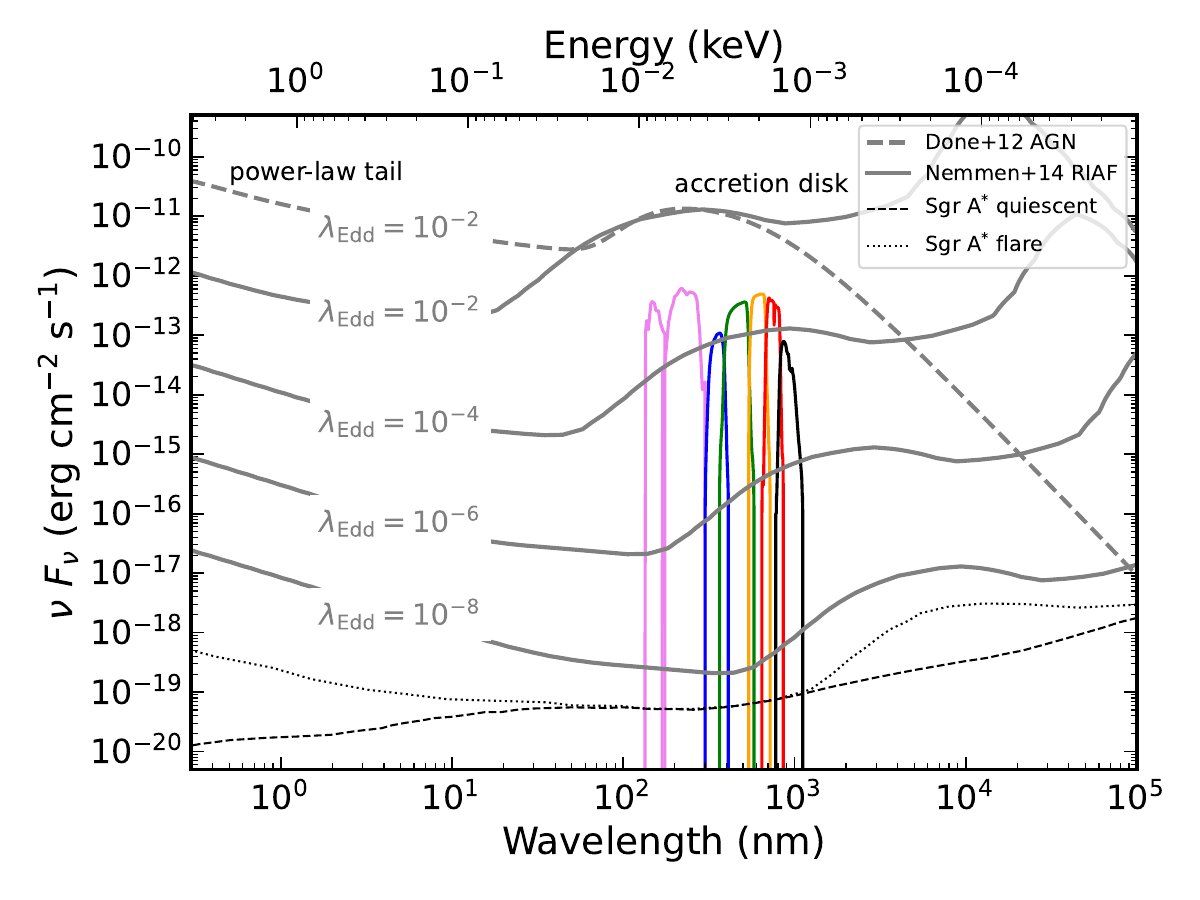}
\caption{Template RIAF spectral energy distributions (SEDs) of AGNs with Eddington luminosity ratios in the range $\lambda_{\rm{Edd}} \equiv L_{\rm{bol}}/L_{\rm{Edd}} = 10^{-8} - 10^{-2}$, with $M_{\rm{BH}} = 4 \times 10^6\ M_{\odot}$ using the model of \citet{Yuan2003} as implemented by \citet{Nemmen2014} (thick gray lines, denoted ``Nemmen+14 RIAF''). For comparison, we show the radiatively efficient accretion model of \citet{Done2012} using the same parameters in Figure~\ref{fig:sed}, except we set the electron temperature for the soft Comptonisation component to $kT_{e} = 1.9$ keV to match Sgr A$^*$ \citep{Baganoff2003} (thick dashed gray lines, denoted ``Done+12 AGN''). We assume a distance of $30$ Mpc for these models. We also show the filter transmittance (throughput) curves for the GALEX (FUV and NUF; violet) and SDSS bandpasses ($ugriz$; blue to black) for reference (arbitrary $y$-axis scaling). We label the approximate locations of the dominant SED component in black text (but note the shift of their peak wavelengths to the right as $\lambda_{\rm{Edd}}$ decreases). The dashed (dotted) black line is the best-fit quiescent-state (flaring-state) \citet{Yuan2003} radiatively inefficient accretion flow disk model for Sgr A$^*$ ($\lambda_{\rm{Edd}} \sim 10^{-8.5}$; $M_{\rm{BH}}=4.3 \pm 0.2 \times 10^{6}\ M_{\odot}$; \citealt{Genzel2010}), assuming a distance of 30 Mpc. Note the \citet{Yuan2003} SED contains outflow/jet synchrotron low-frequency radio emission, while the \citet{Done2012} \textsc{xspec} SEDs do not contain the outflow/jet synchrotron emission. \label{fig:sed2}}
\end{figure*}

\begin{table*}
\caption{Format of the FITS file containing the pre-computed grid of \citet{Done2012} or \citet{Nemmen2014} model SEDs.
\label{tab:sed}}
\small 
\begin{tabular}{lllll}
\hline \hline
Header & Column Name & Format & Unit & Description \\
\hline
0 & data & $^{a}$float64 & $\log_{10}( {\rm erg}\ {\rm cm}^{-2} \ \ {\rm s}^{-1} )$ & $\log_{10}$ of the SED computed on the grid \\
\hline
1 & data & $^{b}$float64 & AB mag & Absolute magnitude in the $i$ band at $z=2$ computed on the grid \\
\hline
2 & log\_M\_BH & float64 & $\log_{10}( M_{\odot} )$ & $\log_{10}$ of the black hole mass \\
2 & log\_LAMBDA\_EDD & float64 & $\log_{10}( M_{\odot} )$ & $\log_{10}$ of the Eddington ratio \\
2 & Z & float64 &  & Redshift \\
\hline
3 & $^{c}$log\_WAV & float64 & $\log_{10}( {\rm nm} )$ & $\log_{10}$ of the rest-frame wavelengths where SED is evaluated \\
\hline
\end{tabular}\\
{\raggedright 
$^{a}$ This is a 4-dimensional array of the shape [log\_M\_BH, log\_LAMBDA\_EDD, Z, log\_WAV]. \\
$^{b}$ This is a 2-dimensional array of the shape [log\_M\_BH, log\_LAMBDA\_EDD]. \\
$^{c}$The wavelength range over which the SEDs are evaluated is $10^{-3} - 10^8$ nm spaced evenly in $\log$ space. \\
\par}
\end{table*}

In order to predict the observed (time-averaged) luminosity in a given band $L_{\rm{band}}$, we need to assume a bolometric correction factor, defined as ${\rm{BC}}_{\rm{band}} = L_{\rm{bol}}/L_{\rm{band}}$. Typically, bolometric corrections are inferred from a template quasar spectral energy distribution (SED). However, the disk temperature profile of an IMBH is expected to differ significantly from that of a SMBH accreting at the same Eddington ratio, causing the SED to peak in the extreme UV (e.g., \citealt{Cann2018}). For this reason, it is inappropriate to use standard AGN or quasar SEDs to explore the IMBH regime (e.g., \citealt{Richards2006}). Instead, here we adopt the energetically self-consistent model of \citet{Done2012} that assumes that the emission thermalizes to a color-temperature-corrected blackbody only at large radii for radiatively efficient accretion ($L_{\rm{bol}}/L_{\rm{Edd}} > 10^{-3}$). This model captures the major components observed in the rest-frame UV/optical in narrow-line Seyfert 1 galaxy SEDs: black-body emission from the outer color-temperature-corrected accretion disk; an inverse Compton scattering of photons from the inner disk model of the soft X-ray excess, and inverse Compton scattering in a corona to produce the power-law tail.

For mass accretion rates $\dot{m}<\dot{m}_{\rm{crit}} \approx \alpha^2 \approx 0.1$, a radiatively inefficient accretion flow (RIAF) is expected to develop, resulting in a much lower luminosity \citep{Fabian1995,Narayan1994,Narayan1995}. It is thought that black holes with $10^{-6}<\dot{m}<\dot{m}_{\rm{crit}}$ may fall in a hybrid RIAF regime, while ``quiescent'' BH with $\dot{m}<10^{-6}$ are in a RIAF-dominated regime \citep{Ho2009}, resulting in a power-law SED like the quiescent-state of Sgr A$^\ast$ \citep{Narayan1998}. The dimensionless mass accretion rate is given by:
\begin{eqnarray}
\dot{m} \simeq 0.7\ (\alpha/0.3)\  (L_{\rm{bol}}/L_{\rm{Edd}})^{1/2},
\end{eqnarray}
where $\alpha$ is the \citet{Shakura1973} viscosity parameter. For RIAFs where, $L_{\rm{bol}}/L_{\rm{Edd}} < 10^{-3}$, we adopt the model of \citet{Nemmen2014}. The model includes an inner advection-dominated accretion flow (ADAF), and an outer truncated thin accretion disk and a jet \citep{Nemmen2014,Yuan2007,Yuan2005}. This model provides a reasonable description for low luminosity AGNs and low-ionization nuclear emission-line region (LINER; \citealt{Eracleous2010,Molina2018}) galaxies with low accretion rates ($L_{\rm{bol}}/L_{\rm{Edd}} \sim 10^{-6} - 10^{-4}$; \citealt{Nemmen2014}). Therefore, we adopt $L_{\rm{bol}}/L_{\rm{Edd}} = 10^{-3}$ as the boundary between radiatively efficient and inefficient accretion flow SEDs, although precisely where this boundary lies is unclear (e.g., \citealt{Ho2009}). 

\subsubsection{Radiatively Efficient Accretion}

To derive bolometric corrections, we use version 12.12.0 of the \textsc{xspec} software\footnote{\url{https://heasarc.gsfc.nasa.gov/xanadu/xspec/}} \citep{Arnaud1996} to generate a fine grid of \citet{Done2012} \texttt{optxagnf} SED models spanning $M_{\rm{BH}} = 10^2 - 10^9\ M_{\odot}$, $L_{\rm{bol}}/L_{\rm{Edd}} = 10^{-3} - 1$, and $z = z_{\rm{min}} - z_{\rm{max}}$. We make the following simple assumptions for the additional parameters in the model: BH spin $a_\star = 0$; coronal radius of transition between black-body emission to a Comptonised spectrum $r_{\rm{cor}} = 100\ R_g$; electron temperature of the soft Comptonisation component (soft X-ray excess) $kT_e = 0.23$ keV; optical depth of the soft excess $\tau=11$; spectral index of the hard Comptonisation component $\Gamma = 2.2$; and fraction of the power below $r_{\rm{cor}}$ which is emitted in the hard Comptonisation component $f_{\rm{pl}} = 0.05$. The outer radius of the disk is set to the self gravity radius \citep{Laor1989}. These parameters are chosen to roughly match that of narrow-line Seyfert 1 galaxy RE1034+396 (see \citet{Done2012} for a more complete description of each parameter). We interpolate this grid of SEDs at each $N_{\rm{draw, BH}}$ Eddington ratio, BH mass, and redshift using our Monte Carlo model. We provide this grid of pre-calculated SEDs as a supporting fits data file\footnote{\url{https://doi.org/10.5281/zenodo.6812008}}. The format of the data file is described in Table~\ref{tab:sed}. We assume no dust extinction/reddening, because the LSST Rubin wide-fast-deep survey is expected to largely avoid the galactic plane and the intrinsic dust extinction in Type 1 AGNs is generally small. Finally, we use the optical filter transmission curves and the SED to compute $L_{\rm{band}}$. The \citet{Done2012} SED models are undefined for $M_{\rm{BH}} > 10^9\ M_{\odot}$ in \textsc{xspec}, so we caution that our derived luminosities for the most massive SMBHs relies on extrapolation from this grid of parameters. Nevertheless, we will show that our derived $L_{\rm{band}}$ values are close to the observed $L_{\rm{band}}$ values from SDSS quasars below.

We show \citet{Done2012} model SEDs spanning $M_{\rm{BH}} = 10^2 - 10^8\ M_{\odot}$ with $L_{\rm{bol}}/L_{\rm{Edd}} = 0.1$ in Figure~\ref{fig:sed}. Our model SEDs are evaluated on a grid spanning $M_{\rm{BH}} = 10^2 - 10^9\ M_{\odot}$, but we have only shown a subset of the results to avoid crowding the figure. We over-plot the SDSS optical \citep{Blanton2017} and GALEX UV \citep{Martin2005} filter transmission curves for reference. For comparison, we also show the best-fit \citet{Gierlinski2009} irradiated disk model of the IMBH candidate HLX-1 \citep{Farrell2009} fit to \emph{Hubble Space Telescope} and \emph{Swift} photometry from \citet{Farrell2014}. This SED model displays qualitatively similar features to the \citet{Done2012} models, given its expected mass of $M_{\rm{BH}} \sim 10^4\ M_{\odot}$ and distance of 95 Mpc \citep{Farrell2014}. Other phenomenological models might also adequately describe the SED arising from an accretion-disk around an IMBH (e.g., \citealt{Mitsuda1984,Makishima1986}). Indeed the SED from the accretion disk emission may differ if the IMBH is in a binary configuration that undergoes state transitions similar to X-ray binaries \citep{Servillat2011}. Here, we assume an IMBH is in a ``high-soft''/rapidly-accreting state where its disk may be approximately geometrically thin and behave like a scaled-down accretion disk around a SMBH \citep{McHardy2006,Scaringi2015,Burke2021c}. One could also incorporate variations in model parameters into our Monte Carlo framework. Although our results depend on these model assumptions, it is unlikely to change our final results in excess of the fiducial uncertainty on the BH mass/bolometric luminosity function. Nevertheless, we retain the flexibility in our framework to substitute other SED models as better observational constraints on dwarf AGN SEDs become available in the future.

\subsubsection{Radiatively Inefficient Accretion}

We calculate \citet{Nemmen2014} RIAF model SEDs\footnote{\url{https://github.com/rsnemmen/riaf-sed}} and add them to our grid of model SEDs spanning $M_{\rm{BH}} = 10^2 - 10^9\ M_{\odot}$, $L_{\rm{bol}}/L_{\rm{Edd}} = 10^{-8} - 10^{-3}$, and $z = z_{\rm{min}} - z_{\rm{max}}$. We make the following
simple assumptions for the additional parameters in the model: power-law index for accretion rate (or density) radial variation $s=0.3$, \citet{Shakura1973} viscosity parameter $\alpha=0.3$, ratio between the gas pressure and total pressure $\beta = 0.9$, strength of wind $p=2.3$, fraction of energy dissipated via turbulence that directly heats electrons $\delta=10^{-3}$, adiabatic index $\gamma=1.5$. The outer radius of the disc is set to the self gravity radius \citep{Laor1989}. These parameters are chosen to roughly match those inferred from fitting a sample of LINERs from \citet{Nemmen2014} (see the Nemmen et al. paper for a more complete description of each parameter). To overcome sensitivities to boundary conditions when finding model solutions, we generate a single template SED with Sgr A$^\ast$-like parameters and normalize the resulting SED by BH mass and accretion rate. We then include the simple color-temperature correction analogous to \citet{Done2012}.

We show \citet{Nemmen2014} RIAF SEDs spanning $L_{\rm{bol}}/L_{\rm{Edd}} = 10^{-8} - 10^{-2}$ along with a \citet{Done2012} SED with $L_{\rm{bol}}/L_{\rm{Edd}} = 10^{-2}$ with $M_{\rm{BH}} = 4 \times 10^6\ M_{\odot}$ in Figure~\ref{fig:sed2} and $kT_{e} = 1.9$ keV to match Sgr A$^*$ \citep{Baganoff2003}. For comparison, we show the SED of Sgr A$^\ast$ ($\lambda_{\rm{Edd}} \sim 10^{-8.5}$; $M_{\rm{BH}}=4.3 \pm 0.2 \times 10^{6}\ M_{\odot}$; \citealt{Genzel2010}) in both its quiescent and flaring states and using the radiatively inefficient accretion flow disk model of \citet{Yuan2003}. We find the \citet{Nemmen2014} models provide a reasonable approximation to the optical/UV/X-ray emission of the flaring-state SED of Sgr A$^\ast$. The difference in the shape of the SED compared to \citet{Done2012} model SEDs is attributed to differences between radiatively efficient and RIAFs cooled by advection \citep{Narayan1994,Narayan1995}. There are many theoretical uncertainties regarding the nature of RIAFs, owing to a lack of high-quality observations. However, these detailed assumptions will only affect the luminosities for sources with very low accretion rates in our model which fortunately do not dominate the variability-selected samples.

\subsection{Optical variability}\label{sec:var}

\begin{figure}
\centering
\includegraphics[width=0.5\textwidth]{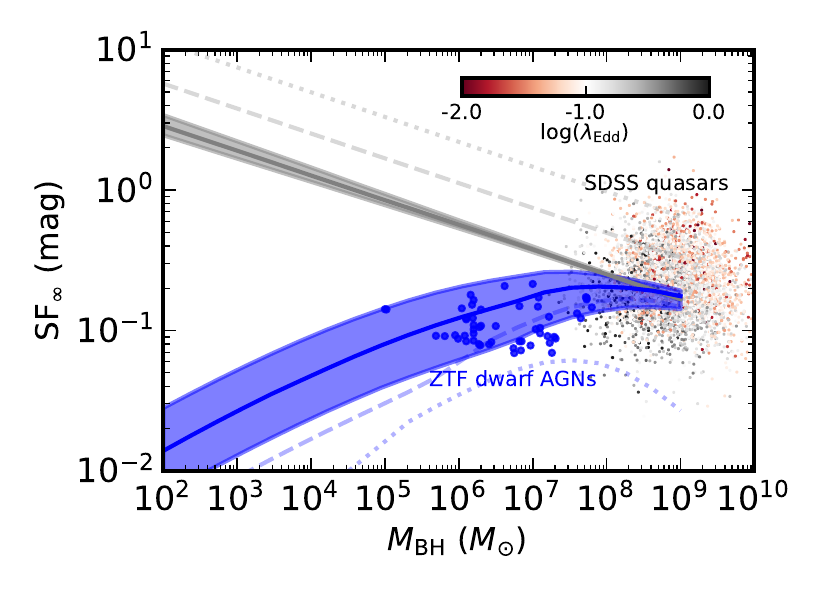}
\caption{Asymptotic rms variability amplitude $\rm{SF}_{\infty}$ versus virial BH mass $M_{\rm{BH}}$ for the sample of SDSS quasars measured from SDSS light curves (points colored by their Eddington luminosity ratios $\lambda_{\rm{Edd}}$; \citealt{MacLeod2010}) and broad-line dwarf AGNs (blue circle symbols) measured from ZTF light curves in the $g$ band computed at $z\sim0.01$. The extrapolated $\rm{SF}_{\infty}$ relations from the  \citet{MacLeod2010} prescription (Equation~\ref{eq:SFinf}) assuming no host galaxy dilution are shown in gray with $L_{\rm{bol}}/L_{\rm{Edd}}=$ 0.1 (solid line), 0.01 (dashed line), and 0.001 (dotted line) with $1\sigma$ uncertainty band shown over the $L_{\rm{bol}}/L_{\rm{Edd}}=$ 0.1 prediction. Our modified extrapolations are similarly shown in blue after accounting for host galaxy dilution assuming a color index of $g-r=0.5$ and a covering factor of $f_\star=10$\%, typical of low-redshift dwarf galaxies. The inconsistency (opposite trends) with the $L_{\rm{bol}}/L_{\rm{Edd}}$ scaling for SDSS quasars and our model is due to the dimming of AGN light as $L_{\rm{bol}}/L_{\rm{Edd}}$ decreases, leading to more host dilution. We have also assumed different host galaxy colors compared to quasar host galaxies (e.g., \citealt{Matsuoka2014}), and the dwarf galaxy and SDSS quasar populations are at different redshifts. The uncertainty is dominated by scatter in the BH-host galaxy relation and the galaxy mass-to-light ratio (see \S\ref{sec:var}). Our modified relation gives more reasonable results in the IMBH regime and is more consistent with observations of dwarf AGN variability. Typical uncertainties on the $\rm{SF_{\infty}}$ measurements are $\sim0.1$ dex. Virial mass uncertainties are typically $\sim0.4$ dex (e.g., \citealt{Shen2013}). \label{fig:SFinf}}
\end{figure}

\begin{figure}
\centering
\includegraphics[width=0.5\textwidth]{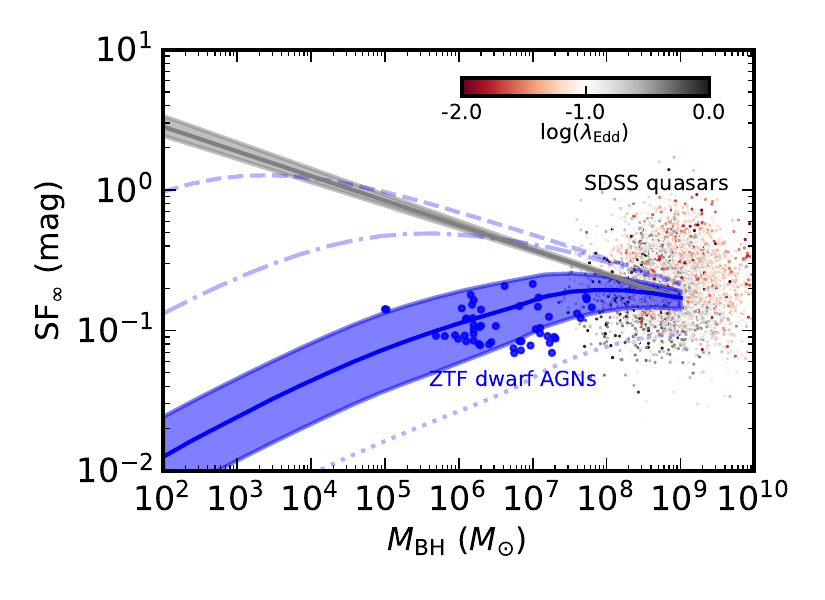}
\caption{Same as Figure~\ref{fig:SFinf} but with $L_{\rm{bol}}/L_{\rm{Edd}}= 0.1$ and varying host dilution covering factors of $f_{\star}= $ 0.2\% (dashed line), 2\% (dash-dotted line), 20\% (solid line), and 100\% (dotted line) with $1\sigma$ uncertainty band shown over the $f_{\star}= $ 20\% prediction. The results with no host dilution are a reasonable approximation of the extrapolated relation for quasars \citep{MacLeod2010}. \label{fig:SFinf2}}
\end{figure}

\begin{figure}
\centering
\includegraphics[width=0.5\textwidth]{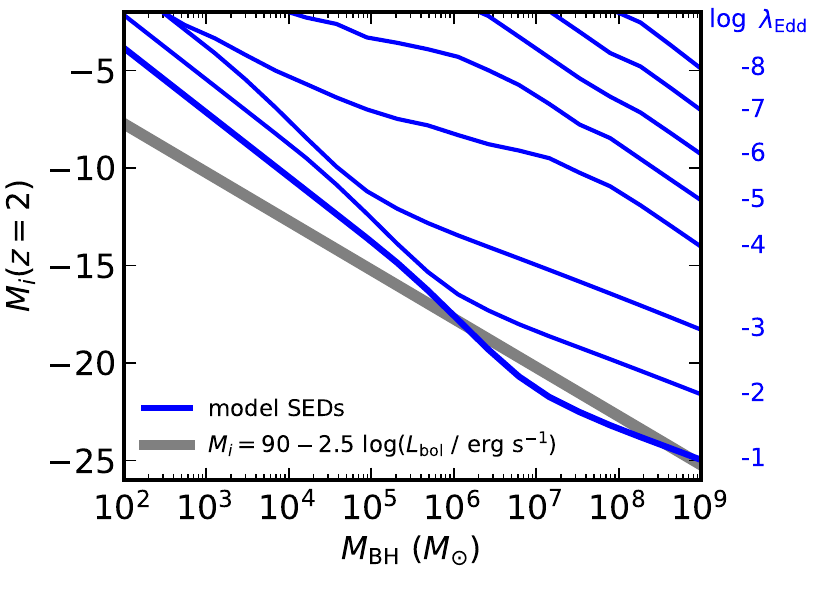}
\caption{Absolute $i$-band magnitude $K$-corrected to $z=2$ versus BH mass computed with the \citet{Done2012} ($L_{\rm{bol}}/L_{\rm{Edd}}>10^{-3}$) or \citet{Nemmen2014} ($L_{\rm{bol}}/L_{\rm{Edd}}<10^{-3}$) SEDs (blue) compared to the relation for quasars $L_{\rm{bol}}/L_{\rm{Edd}}=0.1$ (gray; e.g., \citealt{Shen2009}). The thick blue line is the $L_{\rm{bol}}/L_{\rm{Edd}}=0.1$ case, while the thin blue lines span $L_{\rm{bol}}/L_{\rm{Edd}}=10^{-2} - 10^{-8}$. The width of the gray line corresponds to the $1\sigma$ scatter in the relation. The more complex shape of the blue curve---namely, larger $M_i(z{=}2)$ at lower BH mass---is due to the blueward disk temperature shift at lower BH masses. For $M_{\rm{BH}} < 10^6\ M_{\odot}$, this relation is well-approximated by $M_i = 125 - 3.3\ \log(L_{\rm{bol}}\ /\ {\rm{erg\ s}^{-1}})$. \label{fig:Mi}}
\end{figure}

\begin{figure}
\includegraphics[width=0.5\textwidth]{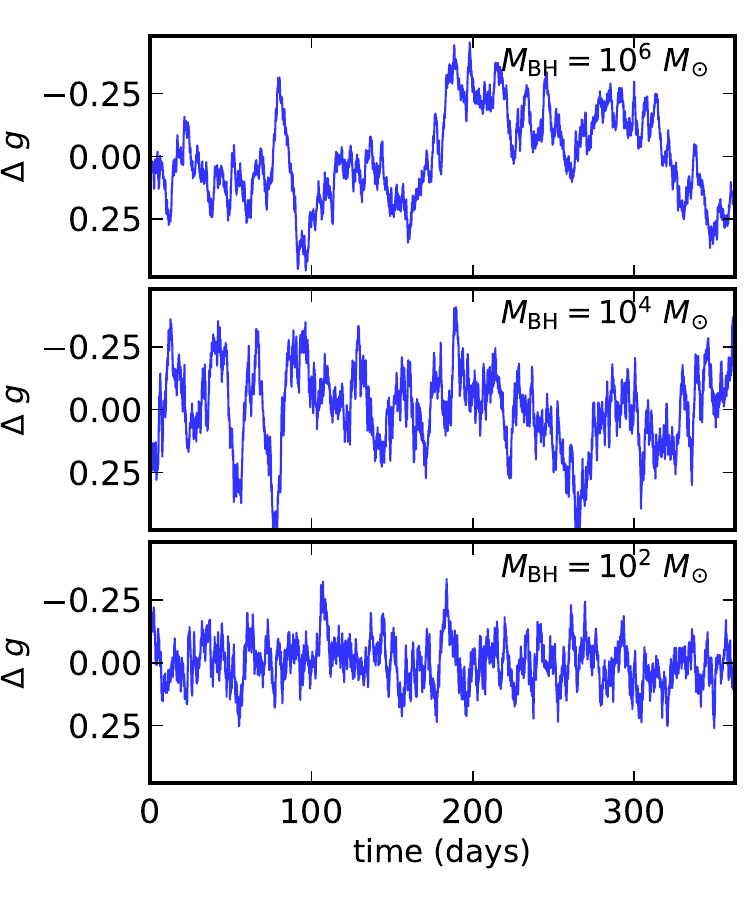}
\caption{Example mock DRW $g$-band rest-frame light curves of AGNs with BH masses in the range $M_{\rm{BH}} = 10^2 - 10^6\ M_{\odot}$, $L_{\rm{bol}}/L_{\rm{Edd}}=0.1$, $g-r=0.5$ and a host dilution covering factor of $f_\star=10$\%, with a duration of 1 year. The mock light curve prescription includes estimates of host galaxy contamination following \S\ref{sec:var}. The variability amplitude of an IMBH saturates at a few tenths of a magnitude due to host dilation and the characteristic timescale of variability is $\sim$ tens of hours.  \label{fig:lc}}
\end{figure}

To a good approximation, AGN light curves can be well described by a damped random walk (DRW) model of variability \citep{Kelly2009,MacLeod2010}. We assume a DRW model for both accretion modes. In the DRW model, the PSD is described by a $f^{-2}$ power-law at the high-frequency end, transitioning to a white noise at the low-frequency end. The transition frequency corresponds to the damping timescale $\tau_{\rm{DRW}}$ as $f_0 = 1/(2\pi\tau_{\rm{DRW}}$). The damping timescale thus describes a characteristic timescale of the optical variability. There is growing evidence that the variability characteristics depend on AGN properties. \citet{Burke2021} found that (i) the damping timescale depends on accretor mass and (ii) there exists a strong correlation between $\tau_{\rm{DRW}}$ and BH mass, which extends to the stellar mass range using optical variability measured for nova-like accreting white dwarfs \citep{Scaringi2015}. We generate mock AGN light curves using the recipe of \citet{MacLeod2010,Suberlak2021}:
\begin{multline}
\label{eq:SFinf}
    \log\left(\frac{{\rm{SF}}_{\infty}}{\rm{mag}}\right) = A + B\ \log\left(\frac{\lambda_{\rm{RF}}}{4000\ {\textup{\AA}}}\right) + 
    C\ (M_i + 23) + \\ D\ \log\left(\frac{M_{\rm{BH}}}{10^9\ M_{\odot}}\right),
\end{multline}
where $A = -0.51 \pm 0.02$, $B = -0.479 \pm 0.005$, $C = 0.131 \pm 0.008$, and $D = 0.18 \pm 0.03$; and,
\begin{multline}
    \log\left(\frac{\tau}{\rm{days}}\right) = A + B\ \log\left(\frac{\lambda_{\rm{RF}}}{4000\ {\textup{\AA}}}\right) + 
    C\ (M_i + 23) + \\ D\ \log\left(\frac{M_{\rm{BH}}}{10^9\ M_{\odot}}\right),
\end{multline}
where ${\rm{SF}}_{\infty}$ is the structure function (SF) evaluated at infinity (i.e., asymptotic rms variability amplitude; e.g., \citealt{Kozlowski2016}) and $A = 2.4 \pm 0.2$, $B = 0.17 \pm 0.02$, $C = 0.03 \pm 0.04$, and $D = 0.21 \pm 0.07$ \citep{Suberlak2021}. Here we adopt the coefficients of $A=2.029\pm0.004$, $D=0.38\pm0.05$ and pivot mass from \citet{Burke2021} which includes dwarf AGNs. In these relations, $\lambda_{\rm{RF}}$ is the rest-frame wavelength of the observation, i.e., $\lambda_{\rm{RF}}=\lambda_{\rm{obs}}/(1 + z)$ where $\lambda_{\rm{obs}}$ is the central wavelength of the filter/band and $z$ is the redshift, and $M_i$ refers to the $i$-band absolute magnitude $K$-corrected to $z=2$, $M_i(z{=}2)$, as a proxy for the AGN bolometric luminosity $L_{\rm{bol}}$ following \citet{Richards2006}. As such, we adopt the relation $M_i = 90 - 2.5\ \log(L_{\rm{bol}}\ /\ {\rm{erg\ s}^{-1}})$ \citep{Shen2009} instead of the actual value computed from the SED (Figure~\ref{fig:Mi}) in these relations so that this variable still acts as a linear proxy for $\log\ L_{\rm{bol}}$ when extrapolated to low BH masses.

We show the predicted $g$-band $\rm{SF}_{\infty}$ versus $M_{\rm{BH}}$ in Figure~\ref{fig:SFinf} using the \citet{Done2012} SEDs to compute $M_i$ (Figure~\ref{fig:Mi}) and varying $L_{\rm{bol}}/L_{\rm{Edd}}=$ 0.1, 0.01, and 0.001. Similarly, we show results for varying host galaxy dilution covering factors of $f_{\star}= $ 0.2\%, 2\%, 20\%, and 100\% in Figure~\ref{fig:SFinf2}. For context, we show the individual data points from SDSS quasars \citep{MacLeod2010} and dwarf AGNs with broad-line (virial) BH mass estimates and $\rm{SF}_{\infty}$ values measured from Zwicky Transient Facility (ZTF; \citealt{Bellm2019}) light curves (Burke et al. in prep). We extrapolate the \citet{MacLeod2010} relation to the IMBH regime, but find the predicted $\rm{SF}_{\infty}$ values of $\gtrsim 1$ mag are far too large to be reasonable. An IMBH with this level of variability has not been detected. The \citet{MacLeod2010} sample is dominated by quasars, so $M_i$ and $\rm{SF}_{\infty}$ correspond primarily to emission from the quasar with a small component contributed by host galaxy. However, in the IMBH regime, host galaxy light is expected to dominate, diluting the variability amplitude from the AGN emission. To estimate this host galaxy light dilution, we use the $M_{\rm{BH}}-M_{\star}$ relation of \citet{Reines2015} (Equation~\ref{eq:MMstar}) and the stellar mass-to-light ratios of \citet{Zibetti2009} assuming a host galaxy color index typical of dwarf AGNs of $g-r \approx 0.5$ (e.g., \citealt{Baldassare2020,Reines2013}) and contamination factor of $f_\star=20$\% (i.e., covering factor, accounting for aperture effects) such that the host galaxy luminosity enclosed in an aperture is $L_{\star, {\rm{ap}}} = f_\star\ L_{\star}$, where $L_{\star}$ is the total luminosity from the host galaxy starlight. These assumptions are justified further in Appendix~\ref{sec:host}, and we will use these mass-dependent parameterizations of the color index and covering factor in our final model. The resulting observed (diluted) rms variability amplitude is,
\begin{equation}
\label{eq:SF1}
    {\rm{SF}}_\infty^{\prime} = \frac{L_{\rm{AGN}}}{L_{\rm{AGN}}+ f_\star L_{\star}}\ {\rm{SF}}_\infty,
\end{equation}
where $L_{\rm{AGN}}$ is the mean AGN luminosity (assumed to be a point source), $L_{\star}$ is the host galaxy luminosity in a given band, and $\rm{SF}_\infty$ is given by Equation~\ref{eq:SFinf}. 

We caution that the assumptions above are highly uncertain (e.g., $\sim0.5$ dex scatter in the $M_{\rm{BH}}-M_{\star}$ relation and $\sim0.3$ dex scatter in the mass-to-light ratios) and the level of host contamination would depend on the individual galaxy. Nevertheless, these qualitative arguments yield more reasonable predictions for the variability amplitude in the IMBH regime and are surprisingly consistent with observations of dwarf AGN variability which have typical $\rm{SF}_\infty$ values of a few tenths of a magnitude \citep{Baldassare2018,Baldassare2020,Burke2020b,Ward2020,Martinez-Palomera2020}. Our modified relation also gives a reasonable prediction for low Eddington ratio black holes. When the AGN emission dominates, the observed anti-correlation between Eddington ratio and variability amplitude (e.g., \citealt{Wilhite2008,Simm2016,Caplar2017,Rumbaugh2018}) may hold for quasars ($M_{\rm{BH}} \sim 10^8 - 10^{10}\ M_{\odot}$; $L_{\rm{bol}}/L_{\rm{Edd}} \sim 0.1$), but below a certain Eddington ratio, the host galaxy dilution becomes so large as to swamp the AGN variability entirely. This is consistent with the lack of detected strong optical variability in very low luminosity AGNs (e.g., detected by ultra deep radio or X-ray surveys) due to host dilution.

We show sample mock DRW $g$-band light curves of AGNs (including host dilution following the prescription above) with BH masses in the range $M_{\rm{BH}} = 10^2 - 10^8\ M_{\odot}$ with $L_{\rm{bol}}/L_{\rm{Edd}} \sim 0.1$ in Figure~\ref{fig:lc} with the same assumptions as above. This figure demonstrates the dramatically more rapid variability ($\lesssim$ days) shown by AGNs in the IMBH regime and suppressed variability amplitude due to estimated host dilution. We compute full mock DRW light curves for all the $N_{\rm{draw, BH}}$ sources in our Monte Carlo model and adopt a simple stellar mass-dependent $g-r$ color index and redshift-dependent contamination factor based on a fitting to SDSS NASA Sloan Atlas galaxies as described in the Appendix~\ref{sec:host}. We assume the emission from the stellar mass of the host star clusters of the wanderers in scenario (iii) are unresolved. This is consistent with the typical size of young star clusters in the local Universe of a few pc or less \citep{Carlson2001}.

\subsection{Optical Type 1 fraction}

\begin{figure}
\centering
\includegraphics[width=0.5\textwidth]{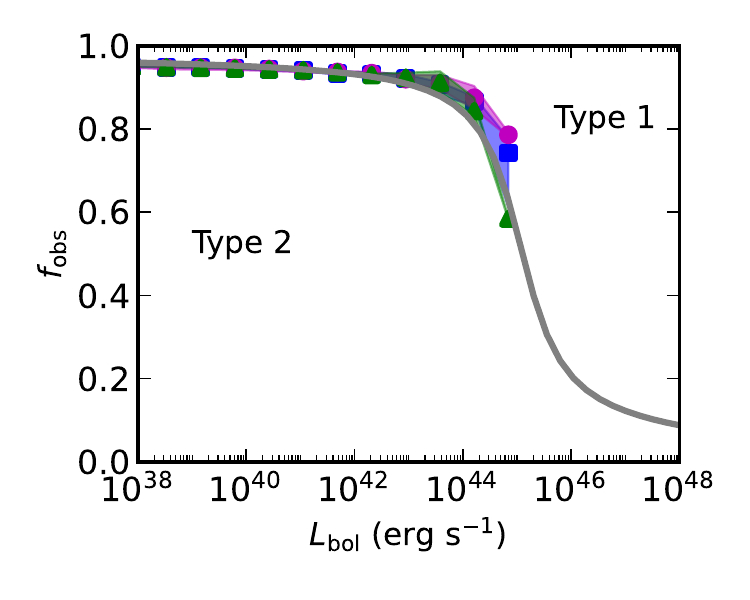}
\caption{Fraction of optically obscured AGNs $f_{\rm{obs}}$ as a function of bolometric luminosity. The gray line shows the model of \citet{Merloni2014} using the X-ray bolometric correction of \citet{Duras2020}. The colored points and $1\sigma$ uncertainty bands are shown for the input ``light'' (blue/square symbols), ``heavy'' (magenta/circle symbols), and ``light $+$ wanderers'' (green triangle symbols) seeding scenarios probed for our LSST-like model. \label{fig:obs}}
\end{figure}

Type 2 (highly optically obscured) AGNs show little or no detectable optical variability because their UV/optical accretion disk emission is thought to be obscured \citep{Barth2014}. We adopt the luminosity-dependent optically obscured AGN fraction $f_{\rm{obs}}$ from \citet{Merloni2014}:
\begin{equation}
    f_{\rm{obs}}(l_x) = A + \frac{1}{\pi} \tan^{-1}\left(\frac{l_0 - l_x}{\sigma_x}\right),
\end{equation}
where $l_x = \log(L_{X} / {\rm{erg}\ {s}^{-1}})$ and their best-fit parameters from their X-ray selected sample are $A=0.56$, $l_0 = 43.89$, an $\sigma_x=0.46$. However, we adopt the normalization $A=0.5$ to ensure $f_{\rm{obs}}$ asymptotes to unity at low luminosity. Formal uncertainties are not given by \citet{Merloni2014}, but the uncertainties in their luminosity bins are $\sim0.2$ dex in luminosity. We show the optically-obscured fraction as function of $L_{\rm{bol}}$ in Figure~\ref{fig:obs} using the luminosity-dependent $2{-}10$ keV bolometric correction of \citet{Duras2020}. We randomly assign each $N_{\rm{draw, BH}}$ sources in our Monte Carlo model to be optically obscured or unobscured using the probability function shown in Figure~\ref{fig:obs}. We simply set the AGN luminosity to zero for optically obscured sources, with Equation~\ref{eq:SF1} ensuring their variability would be undetectable ($\rm{SF}_\infty^\prime \approx 0$ for $L_{\rm{AGN}}/f_{\star}\ L_{\star} << 1$).

\section{Mock Observations}\label{sec:obs}




\subsection{Light curves}

\begin{figure}
\centering
\includegraphics[width=0.5\textwidth]{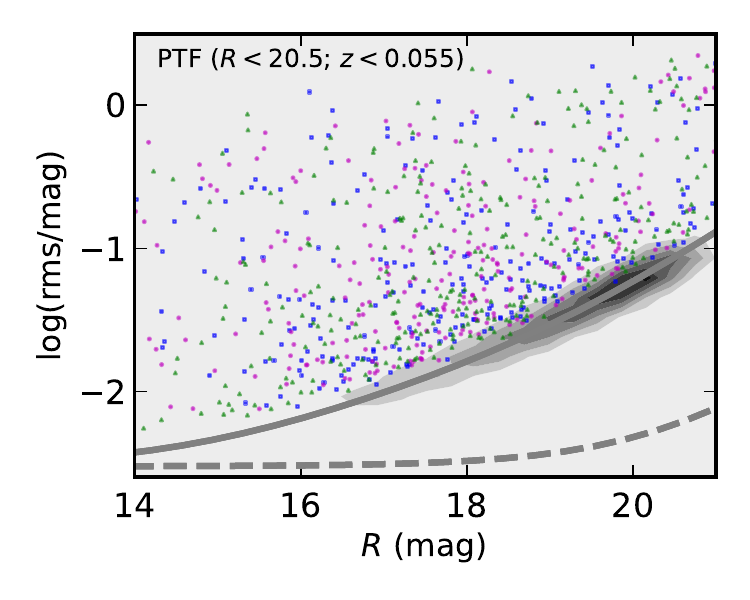}
\includegraphics[width=0.5\textwidth]{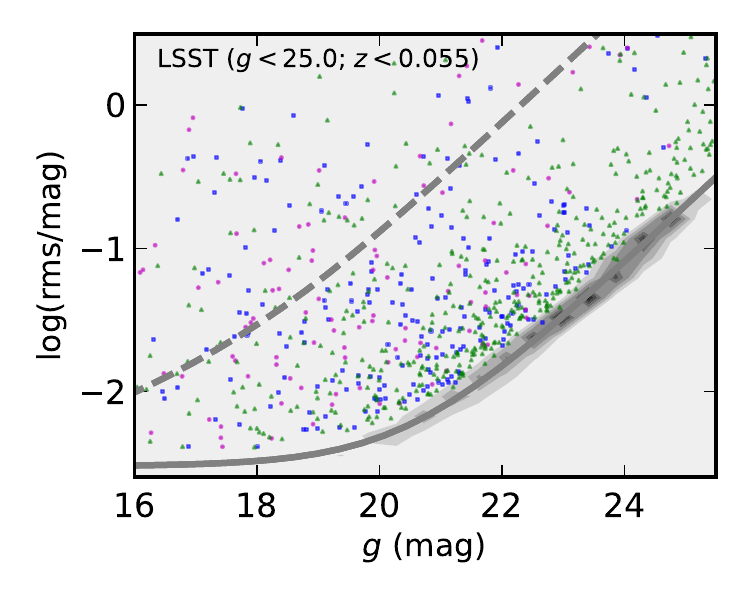}
\caption{Measured light curve rms versus host galaxy aperture magnitude for our PTF (\emph{upper panel}; cf. Figure~3 of \citealt{Baldassare2020}) and LSST-like (\emph{lower panel}) mock samples. The variable sources are shown as colored points (magenta circles: ``heavy'' seed scenario; blue squares: ``light'' seed scenario; green triangles: ``light $+$ wanderers'' scenario), while the shaded contours shows the total light curve distribution (darker contours being regions of higher density). The distributions are from a single, representative bootstrap realization of our model results. The number of data points has been reduced by a factor of 10 (PTF) or 100 (LSST) to improve clarity. The single-visit model photometric precision rms $\sigma_{1}$ versus apparent magnitude for LSST Rubin $g$-band following Equations~\ref{eq:ppm1} and \ref{eq:ppm2} \citep{Ivezic2019} is shown in gray. To facilitate comparison, the dashed lines in the top panel and lower panel show the photometric precision model for LSST Rubin and PTF, respectively. \label{fig:phprec}}
\end{figure}

\begin{figure*}
\includegraphics[width=0.98\textwidth]{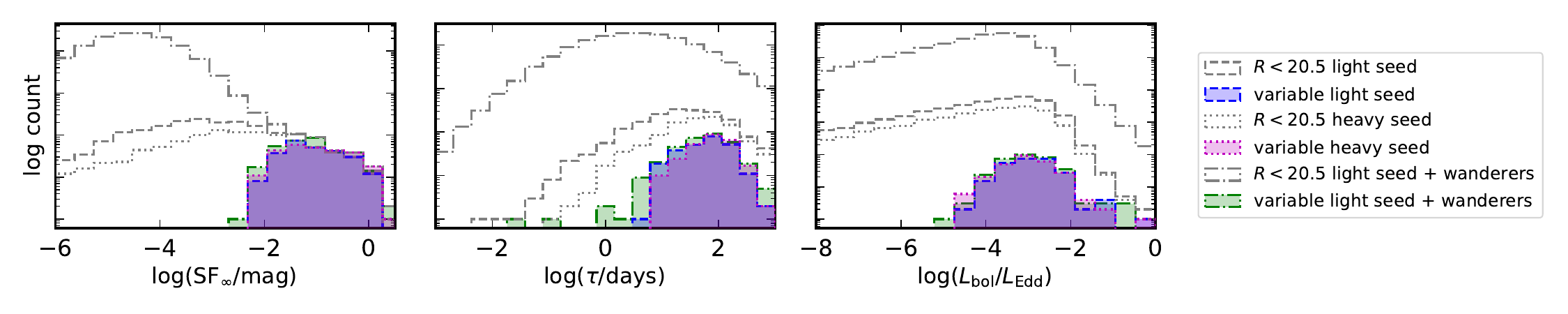}
\includegraphics[width=0.98\textwidth]{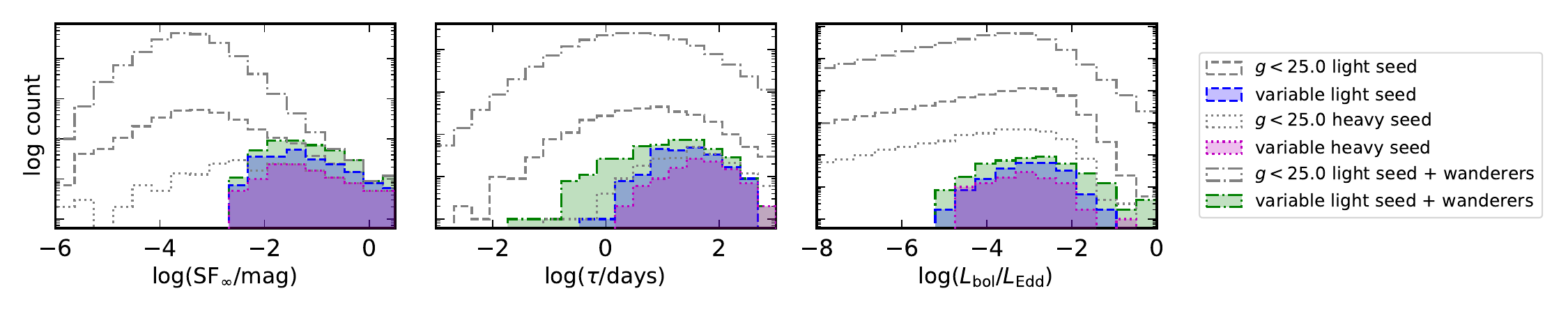}
\caption{Distributions of (host diluted) asymptotic rms variability amplitude ${\rm{SF}_\infty}$, rest-frame damping timescale $\tau$, and Eddington luminosity ratios for all sources within the flux limit of the survey (gray histograms) and variable sources detected in the survey for the different input seeding scenarios (magenta shaded histograms: ``heavy''; blue shaded histograms: ``light''; green shaded histograms: ``light $+$ wanderers'') for our PTF (\emph{upper panel}) and LSST-like (\emph{lower panel}) mock samples. The distributions are from a single, representative bootstrap realization of our model results. This figure demonstrates the resulting distributions of the parameters relative to the input distributions after variability selection.  \label{fig:SFtau}}
\end{figure*}

\begin{figure*}
\centering
\includegraphics[width=0.98\textwidth]{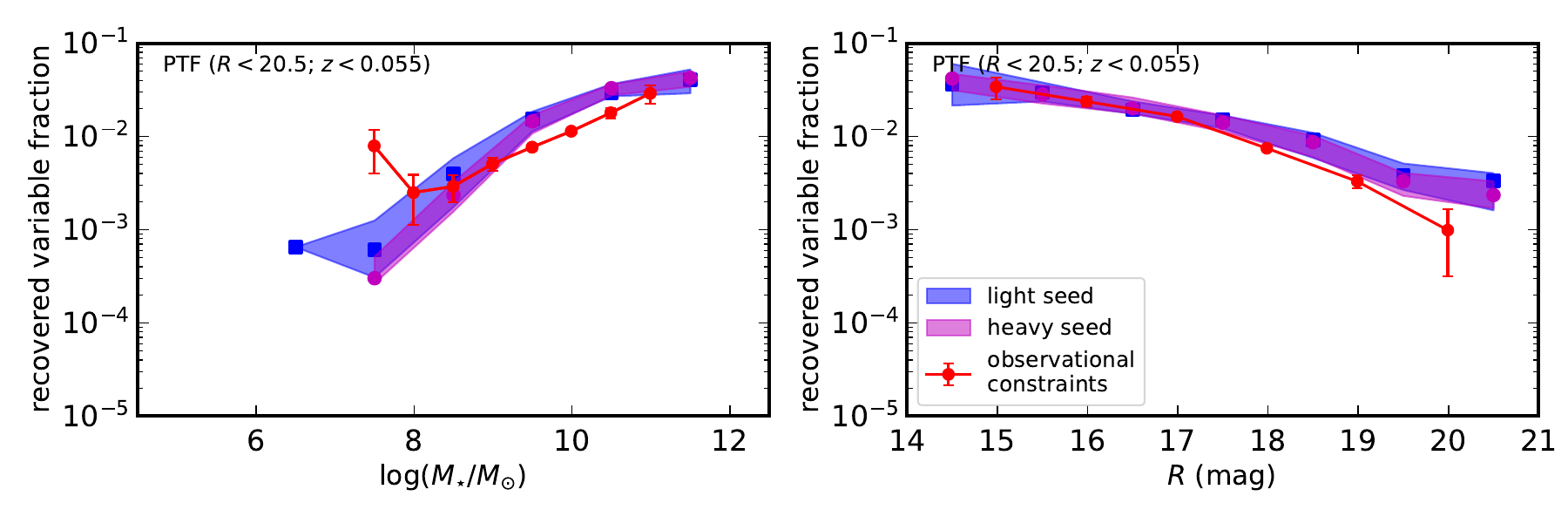}
\includegraphics[width=0.98\textwidth]{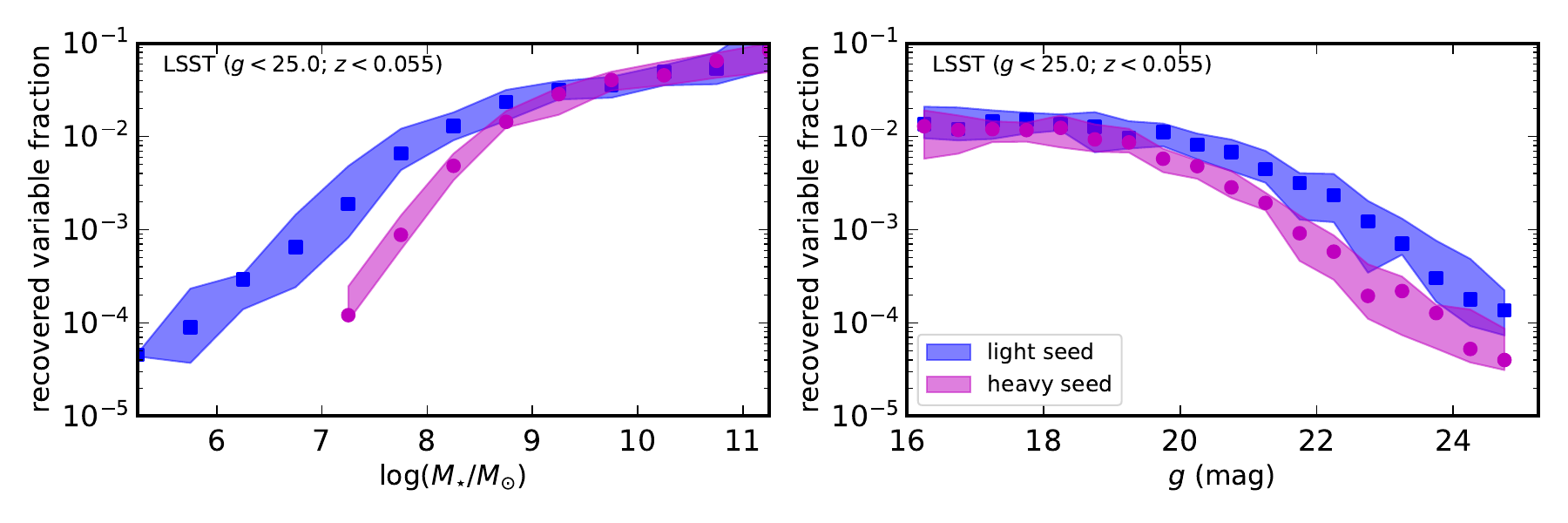}
\caption{Recovered (observed) variable (defined as $\sigma_{\rm{var}} > 2$) AGN fraction versus host galaxy stellar mass (\emph{left}) and aperture apparent magnitude (\emph{right}) for the input ``light'' (blue/square symbols) and ``heavy'' (magenta/circle symbols) seeding scenarios for our PTF (\emph{upper panel}) and LSST-like (\emph{lower panel}) models. These recovered variable fractions are computed by selecting for variable light curves following mock observations as described in \S\ref{sec:obs} after including all components of our demographics model as described in \S\ref{sec:model}. The current observational constraints and $1\sigma$ uncertainties from PTF are shown in red \citep{Baldassare2020}. We omit the data points in the least and most massive bins and faintest bin where the sample is highly incomplete for clarity. Our model results have greater statistical power at low stellar mass than the constraints from \citet{Baldassare2020} because that sample is limited to SDSS spectroscopically-targeted galaxies ($r\lesssim17.8$ mag), which is shallower than the PTF flux limit of $R \sim 20.5$ mag. \label{fig:varfrac}}
\end{figure*}

\begin{figure*}
\centering
\includegraphics[width=0.98\textwidth]{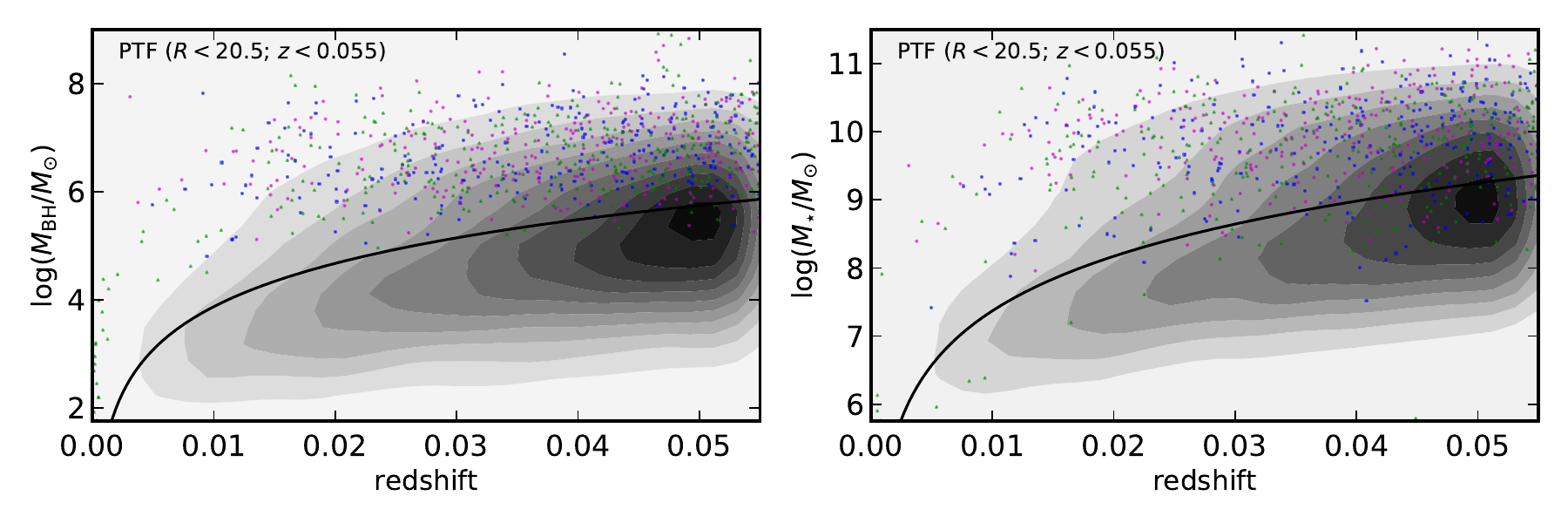}
\includegraphics[width=0.98\textwidth]{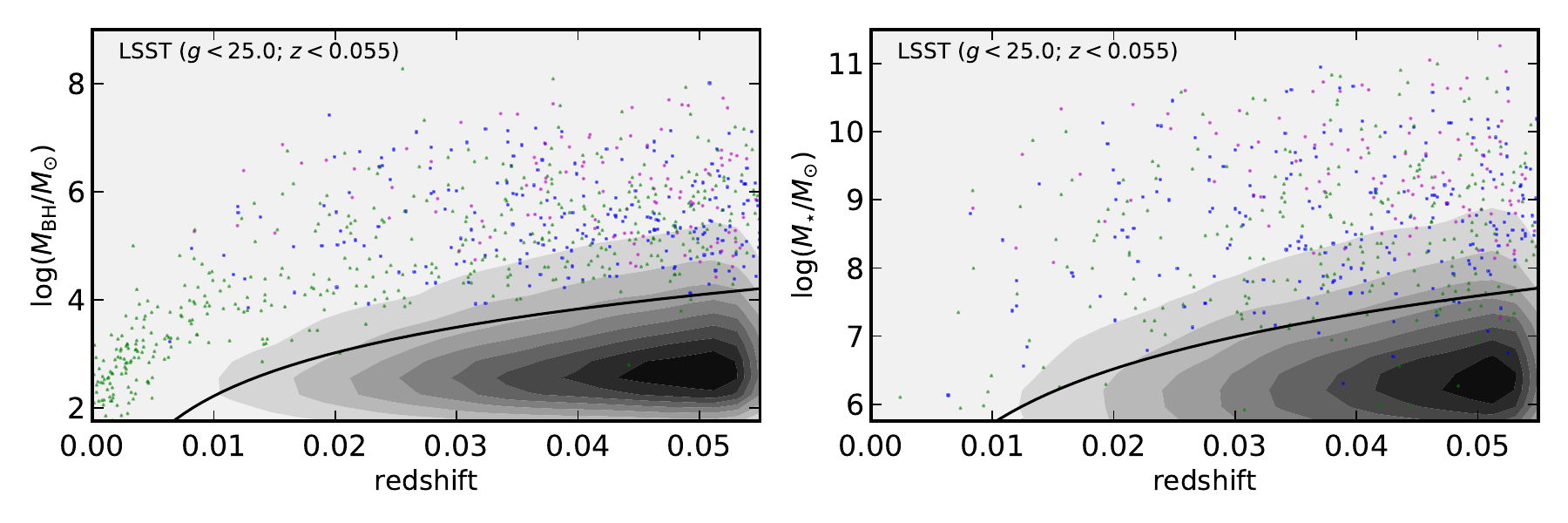}
\caption{BH mass (\emph{left}) and stellar mass (\emph{right}) $-$ redshift distributions for the input ``light'' (blue squares), ``heavy'' (magenta circles), and ``light $+$ wanderers'' (green triangles) seeding scenarios for our PTF (\emph{upper panel}) and LSST-like (\emph{lower panel}) models. We assume measurement uncertainties of $\sim0.3$ dex in stellar mass. Darker contours represent denser regions of the distributions. The scatter points are recovered variable sources are computed by selecting for variable light curves following mock observations as described in \S\ref{sec:obs} after including all components of our demographics model as described in \S\ref{sec:model}. The gray-scale contours represents the underlying distribution of all sources (variable and non-variable) in each model. The solid black curves represents the theoretical mass detection limits following Appendix~\ref{sec:appdetlim} assuming a typical rms variability amplitude of 0.1 mag. The distributions are from a single, representative bootstrap realization of our model results. The number of data points has been reduced by a factor of 10 (PTF) or 100 (LSST) to improve clarity. \label{fig:massredshift}}
\end{figure*}

\begin{figure}
\centering
\includegraphics[width=0.5\textwidth]{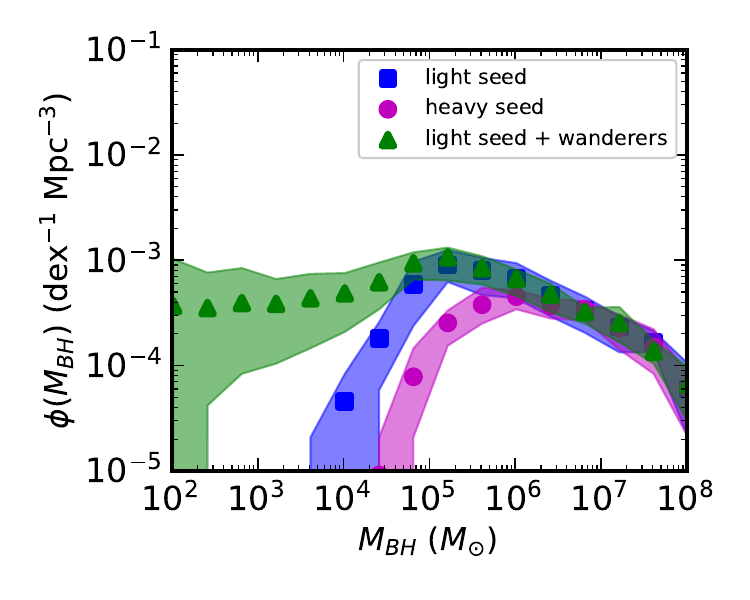}
\caption{Recovered (observed) BH mass function for the input ``light'' (blue/square symbols), ``heavy'' (magenta/circle symbols), and ``light $+$ wanderers'' (green triangle symbols) seeding scenarios for our LSST Rubin-like model. These recovered variable fractions are computed by selecting for variable light curves following mock observations as described in \S\ref{sec:obs} after including all components of our demographics model as described in \S\ref{sec:model}. \label{fig:varfracbh}}
\end{figure}

In order to perform source forecasts, we generate synthetic observations assuming LSST Rubin -like observational parameters. We focus our mock observations on the $g$-band, because the (diluted) AGN variability amplitude is typically larger at bluer wavelengths and the $u$-band suffers from worse single-epoch imaging depth. We generate realistic DRW light curves with a duration of 10 years, a cadence of 25 days, and a season length of 150 days, which roughly matches the expected median values of the ``baseline'' $g$-band LSST Rubin wide-fast-deep survey.\footnote{See \texttt{baseline\_v2.0\_10yrs} metrics at \url{http://astro-lsst-01.astro.washington.edu:8080/allMetricResults?runId=1}} We adopt the photometric precision model of LSST Rubin from \citet{Ivezic2019} of the form:
\begin{equation}
    \label{eq:ppm1}
    \sigma_{1}^2 = \sigma_{\rm{sys}}^2 + \sigma_{\rm{rand}}^2,
\end{equation}
where $\sigma_{1}$ is the expected photometric error in magnitudes for a single visit, $\sigma_{\rm{sys}}$ is the systematic photometric error, and $\sigma_{\rm{rand}}$ is the random photometric error given by, 
\begin{equation}
    \label{eq:ppm2}
    \sigma_{\rm{rand}}^2 = (0.04 - \gamma)\ x + \gamma\ x^2\ {(\rm{mag}^2)},
\end{equation}
with $x\equiv10^{0.4(m-m_5)}$ where $m_{5}$ is the 5$\sigma$ limiting magnitude for point sources in a given band, and $\gamma$ is a band-dependent parameter which depends on sky brightness and instrument properties. We use the expected $g$-band flux limit of $m_{5}=25.0$ mag, $\sigma_{\rm{sys}}^2=0.005$ mag, and $\gamma=0.039$ \citep{Ivezic2019}, which is in good agreement with mock observations from synthetic data \citep{Sanchez2020}. In order to enable comparison with the current observational constraints \citep{Baldassare2020}, we generate similar mock observations with the PTF \citep{Law2009}. We adopt a cadence of 5 days, a season length of 100 days, and a total survey length of 5 years. We use the same photometric precision model from \citet{Ivezic2019} but with an $R$-band flux limit now of $m_{5}=21.5$ mag, $\sigma_{\rm{sys}}^2=0.005$ mag, and $\gamma=0.035$. We obtained these values that approximate the data in Figure~3 of \citet{Baldassare2020} by eye. This is apparently more precise at fixed magnitude than the \citet{Ofek2012} PTF calibration. We show the photometric precision models and measured light curve rms values for LSST Rubin and the PTF in Figure~\ref{fig:phprec}.

Taking our mock light curves with flux-dependent uncertainties, we then use the simple $\chi^2$-based variability metric to compute the variability significance:
\begin{equation}
\left[\chi^2/\nu\right]_{\rm var} = \frac{1}{\nu}\sum^N_{i=1} (m_i-\overline{m})^2 w_i,
\end{equation}
where the weighted mean $\overline{m}$ is given by,
\begin{equation}
\overline{m} = \frac{\sum^N_{i=1}m_i w_i}{\sum^N_{i=1} w_i},
\end{equation}
with weights given by the reciprocal of the squared photo-metric uncertainties $w_i=1/\sigma_i^2$ on each measurement $m_i$ in magnitudes (e.g., \citealt{Butler2011,Choi2014}). We then convert this test statistic to a resulting significance $\sigma_{\rm var}$ in units of $\sigma$. This metric is statistically-motivated, model independent, and fast to compute. Following \citet{Baldassare2020}, we consider a source to be variable if its light curve satisfies $\sigma_{\rm var}>2$, which implies a $\sim5\%$ false positive rate. We require the light curve input rms variability amplitude $\rm{SF}_\infty$ to be larger than the survey's photo-metric precision, i.e., $\rm{SF}_\infty > \sigma_1(m)$, where $m$ is the magnitude of the source and $\sigma_1$ is the photo-metric precision model (Equation~\ref{eq:ppm1}) to assure that our variable sources are reliable detections. Our model does not include other contaminants, such as other variable transients (e.g., supernovae, tidal disruption events, or variable stars), or other (possibly non-Gaussian) systematic sources of light curve variability (i.e., non-photometric observations). Therefore, we have no need to introduce a classification metric for ``AGN-like'' variability. This makes our selection simpler and less dependent on the exact underlying process describing AGN light curves but more idealized than reality. We show histograms of the ${\rm{SF}}_{\infty}$, $\tau$, and $\lambda_{\rm{Edd}}$ values for our sources in Figure~\ref{fig:SFtau}, highlighting our detected variable sources from realistic LSST Rubin -like light curves. 

\subsection{Observational Forecasts}

\begin{table}
\centering
\caption{Number of expected IMBHs and massive BHs detectable with LSST Rubin at $z<0.055$ over the WFD footprint.}
\label{tab:num}
\small
\begin{tabular}{ccc}
\hline \hline
Seeding Scenario & Number IMBHs$^{a}$ & Number massive BHs$^{b}$ \\
\hline
light (i) & $3.9^{+4.1}_{-3.0} \times 10^{2}$ & $1.5^{+0.6}_{-0.6} \times 10^{3}$ \\
heavy (ii) & $5.9^{+5.9}_{-5.9} \times 10^{0}$ & $5.9^{+1.5}_{-1.1} \times 10^{3}$\\
light $+$ wanderers (iii) & $9.7^{+6.2}_{-6.9} \times 10^{3}$ & $2.1^{+0.3}_{-0.7} \times 10^{4}$ \\
\hline
\end{tabular}\\
{\raggedright
$^{a}$ $10^2\ M_{\odot} < M_{\rm{BH}} < 10^4\ M_{\odot}$. \\
$^{b}$ $10^4\ M_{\odot} < M_{\rm{BH}} < 10^6\ M_{\odot}$. \\
\par}
\end{table}

We compute the recovered (observed) fraction of variable galaxies in bins of stellar mass and magnitude using the criteria $\sigma_{\rm var}>2$ for both the LSST Rubin ($g<25.0$ mag) and the PTF ($R<20.5$ mag) in Figure~\ref{fig:varfrac}. We assume a bright saturation limit of $R > 14$ mag for the PTF \citep{Ofek2012} and $g > 16$ mag for LSST Rubin \citep{Ivezic2019}. The uncertainties in the figure trace the uncertainties in the model itself. The slight uptick in the smallest mass bin for the PTF light seed scenario can result from small number statistics, because the smallest bins which only contain a few sources. Recall that we have assumed $z<0.055$ and consider a source to be variable if $\sigma_{\rm var}>2$ and the rms variability is larger than the uncertainty given by the photo-metric precision model.\footnote{ One need not necessarily use the rms constraint when constructing a version of Figure~\ref{fig:varfrac}, although the number of false positive detections would likely increase if this is not done. In fact, the $\sigma_{\rm{var}}$ threshold can be lowered further or a different measure, such as the rolling average $\sigma_{\rm{var}}$ versus stellar mass, could be adopted which may be more sensitive to the input occupation fraction.}. We assume total survey solid angles of $\Omega=9380$ deg$^2$ and $\Omega=18,000$ deg$^2$ for the PTF and LSST Rubin, respectively. We show the distribution of stellar mass versus redshift for a single, representative bootstrap realization of our model results in Figure~\ref{fig:massredshift}.

We also compute the recovered fraction of variable galaxies versus BH mass for our LSST Rubin-like model in Figure~\ref{fig:varfracbh}, albeit the BH mass is not usually a directly observable quantity. Assuming an LSST Rubin-like footprint of $\Omega=18,000$ deg$^2$, the number of expected IMBHs in the mass range $10^2\ M_{\odot} < M_{\rm{BH}} < 10^4\ M_{\odot}$ and ``massive black holes'' $10^4\ M_{\odot} < M_{\rm{BH}} < 10^6\ M_{\odot}$ using optical variability for the various occupation fractions used in this work are enumerated in Table~\ref{tab:num}. Similar figures divided into the blue and red galaxy populations is shown in Appendix~\ref{sec:appredblue}. Our calculations indicate that LSST Rubin may be a very promising source for uncovering massive black holes and IMBH candidates modulo the underlying occupation fraction.

\subsection{Recoverability of black hole masses from variability timescales}
\label{sec:varrecovery}

\begin{figure*}
\centering
\includegraphics[width=0.32\textwidth]{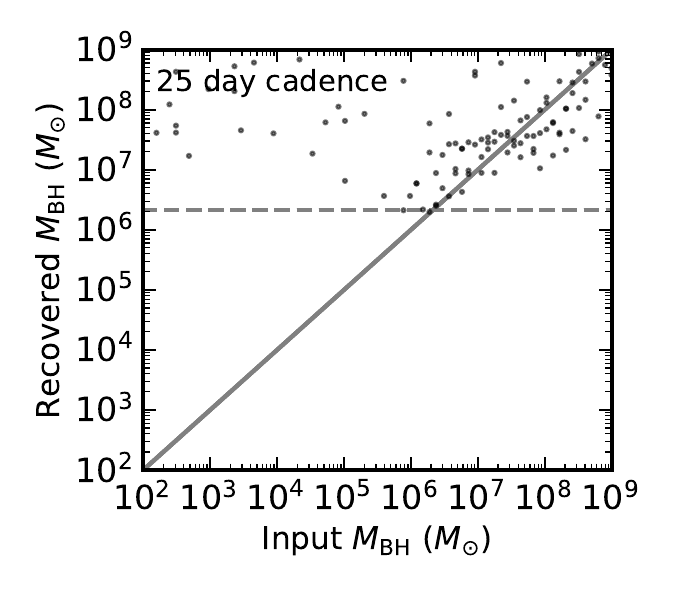}
\includegraphics[width=0.32\textwidth]{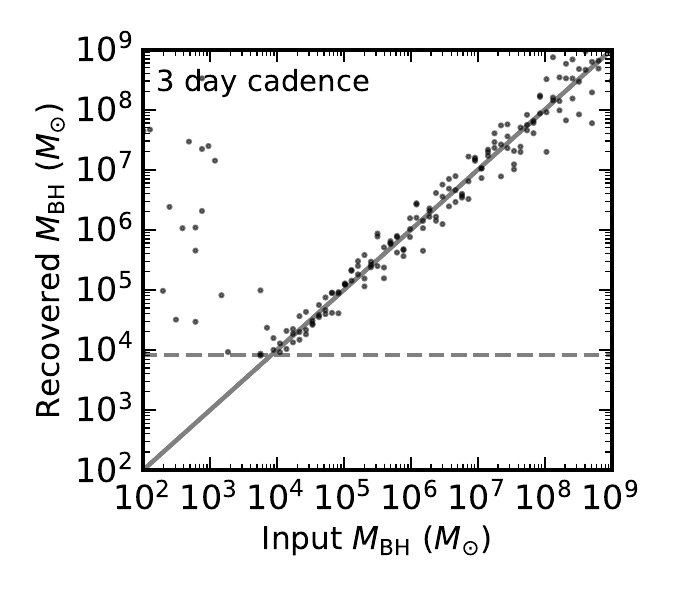}
\includegraphics[width=0.32\textwidth]{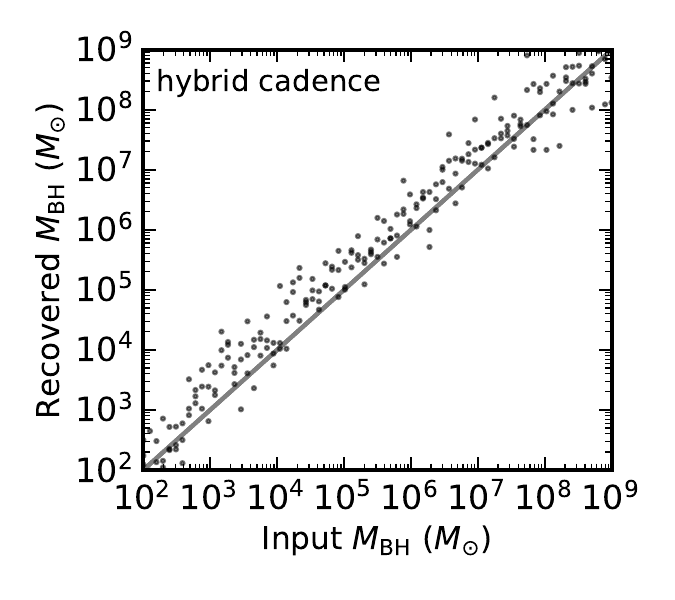}
\caption{Recovered BH mass versus input BH mass for using mock light curves for various cadences scenarios of 3 days, 25 days, and a hybrid cadence of 25 days plus a $\sim$hourly cadence for 5 days assuming the BH mass$-$damping timescale relation of \citet{Burke2021}. The horizontal dashed gray line represents the BH mass with variability timescale equal to the limiting cadence of the light curves. The $y=x$ line is shown as a gray solid line. The hybrid cadence provides the best recovery of IMBH masses measured from realistic light curves of the three cadence modes.  \label{fig:MBHrecovered}}
\end{figure*}

In order to determine how well one can recover the BH mass using optical variability information alone, we attempt to infer the input damping timescale $\tau$ values using mock light curves using different cadence scenarios. Because the dependence of the damping timescale on wavelength is weak \citep{MacLeod2010,Suberlak2021,Stone2022}, observations from multiple bands could be effectively combined to reduce the typical cadence to a few days. Recall that we have used the relation between $\tau$ and BH mass from \citet{Burke2021} to generate the mock DRW light curves. We then use the \textsc{celerite} \citep{Foreman-Mackey2017} package to infer $\tau$ values from these light curves following the procedure of \citet{Burke2021} using a maximum-likelihood fitting of a DRW Gaussian process to the light curve. Deviations from the DRW approximation may complicate the inference of a damping timescale. However, a more sophisticated analysis can be used to measure the damping timescales accurately \citep{Stone2022}. Our resulting recovered BH mass values from optical variability as a function of the input BH masses are shown in Figure~\ref{fig:MBHrecovered} for sources that are significantly variable $\sigma_{\rm{var}}>2$ with an input cadence of 25 days ($g$-band wide-fast-deep cadence), 3 days (wide-fast-deep cadence combining all bands), and a hybrid cadence described below.

Unsurprisingly, we find that we are unable to recover BH mass values below $M_{\rm{BH}} \sim 10^{6.4}\ M_{\odot}$ ($M_{\rm{BH}} \sim 10^{4.1}\ M_{\odot}$) given the limiting input cadence of $25$ ($3$) days. Using the \citet{Burke2021} relation, a $\tau$ value of 25 days corresponds to $M_{\rm{BH}}\sim10^5\ M_{\odot}$ with a $\sim0.3$ dex scatter in the BH mass direction. However, such IMBHs can be identified in principle from their significant variability, and the cadence can be used as a rough upper-limit on the BH mass. We caution that other measures to select AGNs from the auto-correlation information are likely to miss AGNs with characteristic variability timescales less than the survey cadence, because such variability would be nearly indistinguishable from (uncorrelated) white noise. In order to test the feasibility of using a custom designed high-cadence mini-survey to identify IMBHs, we repeat the procedure above using a rapid cadence of observations separated by 2.4 hours for 5 days but with daytime gaps, followed by the standard wide-fast-deep cadence. This hybrid cadence is able to recover the input BH mass values reasonably well, albeit with increased scatter. These relations are derived from a subset of the total AGN population, and the true dependence on other parameters like Eddington ratio as well as the exact cadence adopted.

We have used maximum-likelihood point estimates in Figure~\ref{fig:MBHrecovered} to demonstrate the variability timescale recoverability. This can give results that are slightly systematically offset from the input depending on the input cadence. In this example, the hybrid cadence slightly over-estimates the input BH masses. However, the offset can be mitigated using Markov chain Monte Carlo sampling and to obtain more robust estimates of the input timescales with parameter uncertainties, which are typically $\sim0.5$ dex, and larger than any systematic offset. \citep{Burke2021,Stone2022}.

\section{Discussion} \label{sec:discussion}

\subsection{Comparison with previous Work}

\subsubsection{Variable fraction}

We have constructed a mock sample consistent with the PTF survey ($R<20.5$) to enable direct comparison with observed constraints on the optical variable fraction. We match the sample redshift distribution and survey parameters to observations \citep{Baldassare2020}. Our PTF-like model's recovered variable fraction for all occupation fractions tested here is consistent with \citet{Baldassare2020} within $2\sigma$ below $M_{\star} \sim 10^{9}\ M_{\odot}$. The larger discrepancy at high stellar masses could perhaps be explained by larger contamination in the \citet{Baldassare2020} sample at these masses due to non-AGN variability or some form of incompleteness. For example, more massive AGNs with luminous blue/UV emission could be confused as lower mass star-forming galaxies, flattening out the observed variability fraction. Another obvious possibility is errors from assumptions or extrapolations of uncertain relations in our model. For example, the exact dependence of the derived variability amplitude on the AGN luminosity and accretion rate. The $M_\star \sim 10^{7.5}\ M_{\odot}$ bin shown in Figure~\ref{fig:varfrac} has a $\sim 2\sigma$ discrepancy with our model results. There are only 519 total variable and non-variable sources in that stellar mass bin, and the smallest bin of $M_\star \sim 10^{7.0}\ M_{\odot}$ in Figure~5 of \citet{Baldassare2020} has just 151 total sources (excluded from our Figure~\ref{fig:varfrac}) compared to thousands or tens of thousands of total source in the more massive bins. Therefore, we attribute this fluctuation near $M_\star \sim 10^{7.5}\ M_{\odot}$ to low number statistics. Nevertheless, we consider this agreement to be excellent given the assumptions made in our model.

\subsubsection{Active fraction}

\begin{figure*}
\centering
\includegraphics[width=0.98\textwidth]{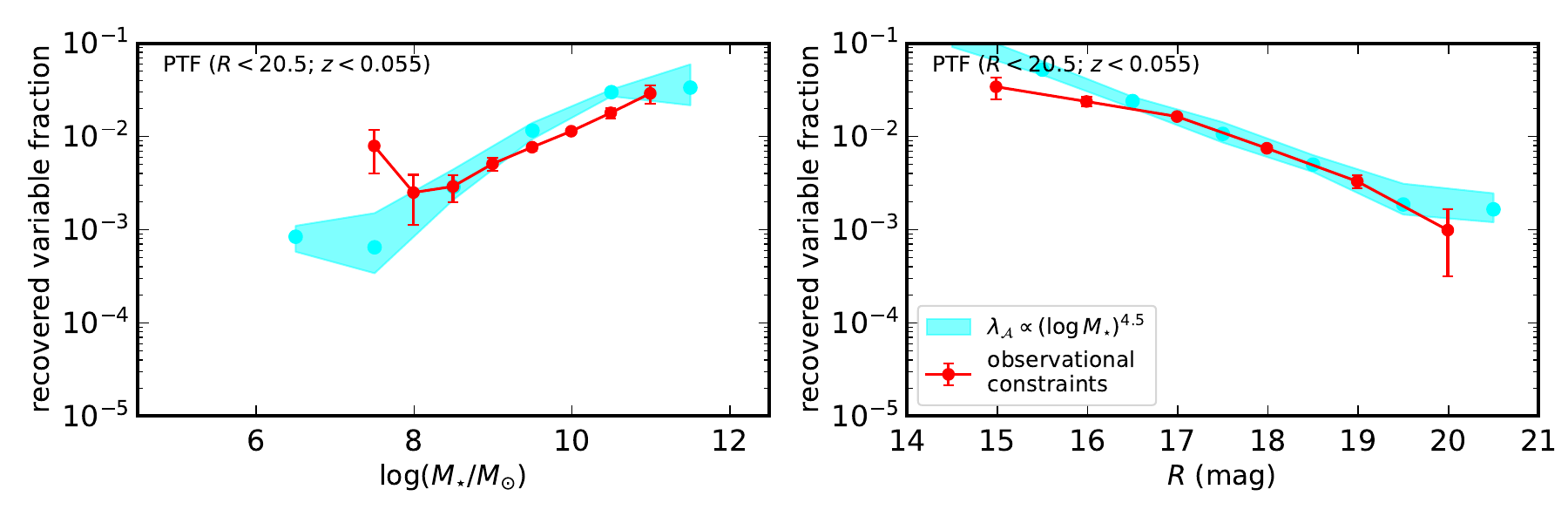}
\caption{Recovered (observed) variable AGN fraction in bins of stellar mass (\emph{left}) and aperture apparent magnitude (\emph{right}) for the active fraction prediction from \citet{Pacucci2021} (cyan/square symbols) for our PTF-like model. An active fraction of the form $\lambda_{\mathcal{A}} \propto (\log{M_{\star}})^{4.5}$ is very similar shape to our model results in Figure~\ref{fig:varfrac} and is a reasonable match to the observational constraints. These recovered variable fractions are computed by selecting for variable light curves following mock observations as described in \S\ref{sec:obs} after including all components of our demographics model as described in \S\ref{sec:model}. The current observational constraints and $1\sigma$ uncertainties from PTF are shown in red \citep{Baldassare2020}. We omit the data points in the most massive and faintest bins where the sample is highly incomplete for clarity. \label{fig:varfracA}}
\end{figure*}

The active fraction---the fraction of galaxies radiating with Eddington luminosity ratio greater than $\lambda_{\rm{Edd, lim}}$---can be defined as,
\begin{equation}
    \lambda_{\mathcal{A}}(M_{\star}, \lambda_{\rm{Edd, lim}}) = \lambda_{\rm{occ}}(M_{\star}) \times\frac{\int^{\lambda_{\rm{Edd, max}}}_{\lambda_{\rm{Edd, lim}}} \xi(\lambda_{\rm{Edd}}, \xi^\ast{=}1)\ d\lambda_{\rm{Edd}}}{\int^{\lambda_{\rm{Edd, max}}}_{\lambda_{\rm{Edd, min}}} \xi(\lambda_{\rm{Edd}}, \xi^\ast{=}1)\ d\lambda_{\rm{Edd}}}
\end{equation}
within the context of our model, where the ERDF $\xi$ is given by Equation~\ref{eq:ERDF}. Our definition differs slightly from the definitions adopted by other authors, who count any galaxy with an assigned $\lambda_{\rm{Edd}}$ value greater than $\lambda_{\rm{Edd, min}}$ toward the active fraction (e.g., \citealt{Weigel2017}). In this work, we have assigned each BH a $\lambda_{\rm{Edd}}$ value, but allow $\lambda_{\rm{Edd}}$ to be so small that the accretion activity effectively goes undetected.


A different approach was adopted by \citet{Pacucci2021}, who developed an alternate theoretical model to predict the active fraction of dwarf AGNs. Their approach derives the active fraction from the number density and angular momentum content of the gas at the Bondi radius (as a proxy for the angular momentum content near an IMBH). After calibrating the model to observations, \citet{Pacucci2021} find an active fraction $\lambda_{\mathcal{A}} \propto (\log{M_{\star}})^{4.5}$ for $10^{7}\ M_{\odot}<M_{\star}<10^{10}\ M_{\odot}$ for black holes accreting at $\lambda_{\rm{Edd}} \sim 0.1$. These arguments imply that the observed optically-variable fraction is roughly the product of the optically unobscured fraction and the active fraction $\lambda_{\rm{var}} \sim (1-f_{\rm{obs}}) \times \lambda_{\mathcal{A}}$.

In our model, we have assumed two mass-independent ERDFs for the blue/green (generally less massive, radiatively efficient accretion) and red (generally more massive, radiatively inefficient accretion) galaxy populations \citep{Weigel2017}. In contrast, the arguments from \citet{Pacucci2021} can be interpreted as a stellar mass dependent ERDF (also see~\citealt{Shankar2013,Hickox2014,Schulze2015,Bongiorno2016,Tucci2017,Bernhard2018,Caplar2018}) as opposed to a galaxy color/type dependent one. To test what impact these different assumptions have on the results, we re-run our forward Monte Carlo model, substituting a continuum of Eddington ratios given by an ERDF for an active fraction of the functional form $\lambda_{\mathcal{A}} = 0.1 \times \left[\log({M_{\star}/M_{\odot}})/9\right]^{4.5}$, which closely matches the normalization in Figure 3 of \citet{Pacucci2021}. Here, active galaxies are assumed to have $\lambda_{\rm{Edd}} = 0.1$ with a dispersion of 0.2 dex (typical for low-$z$ AGN samples; \citealt{Pacucci2021,Greene2007}) and non-active galaxies have $\lambda_{\rm{Edd}} \approx 0$ as determined by random sampling. Our resulting detected variable fraction versus stellar mass for the PTF-like scenario is shown in Figure~\ref{fig:varfracA}.

The resulting variable fraction has a very similar form as our model results. The computed variable fraction has a qualitatively similar scaling with magnitude and mass, which implies that the assumption of a mass-dependent ERDF does not strongly change the results, as expected if radiatively-efficient AGNs dominate the census. This is consistent with the findings of \citet{Weigel2017}. Therefore, we can conclude that our results and the existing observational constraints are broadly consistent with an active fraction of the form $\lambda_{\mathcal{A}} \propto (\log{M_{\star}})^{4.5}$ after calibration to the definition of ``active'' to the level of detectable accretion activity. This is reassuring and points to the fact that our model assumptions are reasonable. However, this simple Gaussian ERDF may not be consistent with the local AGN luminosity function.


\subsection{The effect of uncertainty on stellar mass measurements}\label{sec:syserr}

The broad-band SED of galaxies can be used to infer the stellar mass of galaxies in large photo-metric catalogs. Uncertainties on these stellar masses are typically $\sim 0.3$ dex and dominated by systematic uncertainties from model choices in stellar evolution (e.g., initial mass function, star formation history; \citealt{Ciesla2015,Boquien2019}). An additional problem is degeneracies between star-formation and AGN power-law emission. For example, Type 1 quasars with a blue/UV power-law continuum emission from the accretion disk (i.e., ``big blue bump'') can be confused for dwarf starburst galaxies. This degeneracy can be more problematic when the redshift of the galaxy is uncertain or highly degenerate. Finally, variability from non-simultaneous observations can introduce additional errors in the SED. Because spectroscopic redshifts will not be available for every source in the large planned time-domain surveys, future work is needed to determine the strength of these degeneracies and how they can possibly be minimized (e.g., using the variability amplitude and timescale to independently constrain the strength of the AGN emission) over the entire range of stellar masses. 

We then consider how uncertainties on stellar mass measurements affects the occupation function analysis in Figure~\ref{fig:varfracA}, regardless of the exact sources of the uncertainty. To do this, we repeat the analysis of the variable fraction in Figure~\ref{fig:varfracA}, which assumes a $0.3$ dex uncertainty in stellar mass, using increasingly larger uncertainties of $0.6$ and $0.9$ dex in stellar mass. The results are shown in Appendix~\ref{sec:appsyserr}. We have assumed a Gaussian distribution for the uncertainties, which may not be strictly true. We see that as the uncertainties increase, the variable fraction ``flattens out'' as the stellar masses are smeared into adjacent bins and would result in a larger number of false positive IMBH candidates.

\subsection{Recovery of the host galaxy-black hole mass scaling relation}

\begin{figure}
\centering
\includegraphics[width=0.5\textwidth]{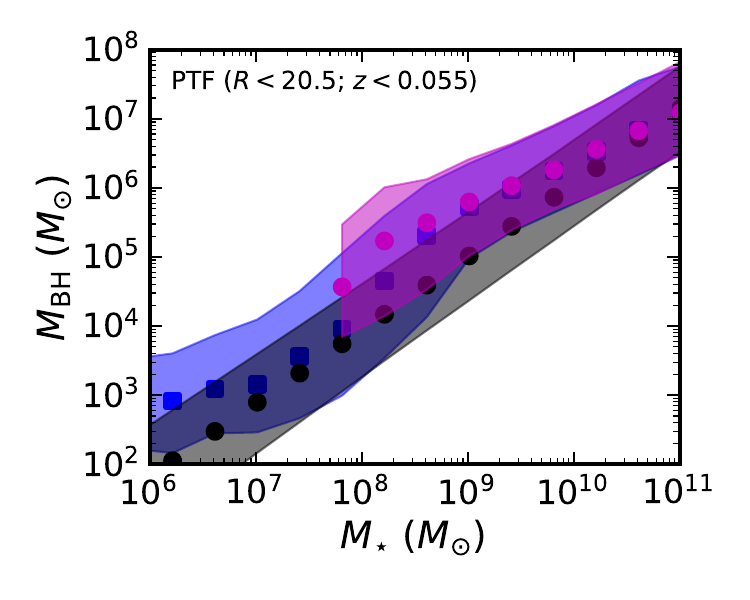}
\includegraphics[width=0.5\textwidth]{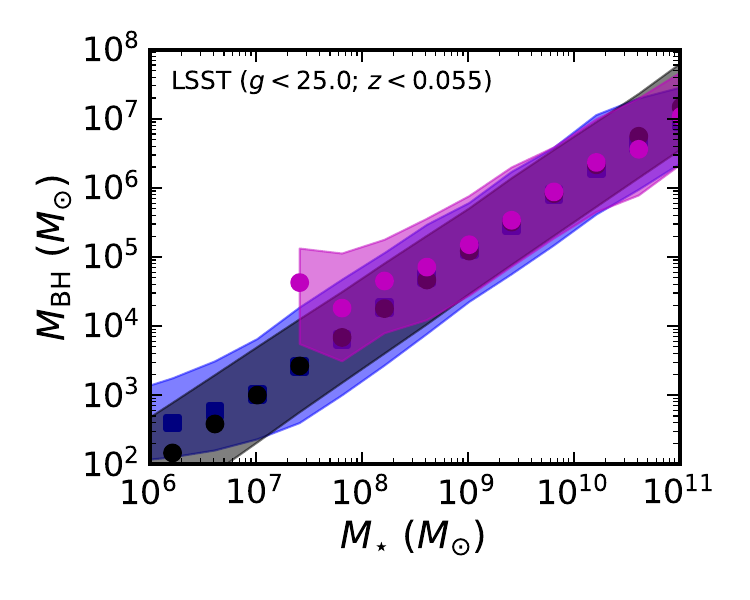}
\caption{Recovered $M_{\rm{BH}}-M_{\star}$ relation for variability-selected sources for the ``heavy'' (blue/square symbols) and ``light'' (magenta/circle symbols) seeding scenarios (strongly overlapping) compared to the input relation given by \citet{Reines2015} (gray) for PTF (\emph{upper panel}) and LSST-like (\emph{lower panel}) models. \label{fig:massbias}}
\end{figure}

We show the recovered $M_{\rm{BH}}-M_{\star}$ relation for variability-selected sources to investigate the influence of variability selection effects in Figure~\ref{fig:massbias}. The more massive and luminous black holes tend to have larger observed variability amplitudes at fixed stellar mass due to having less host galaxy dilution (see discussion in \S\ref{sec:var}). See \citet{Lauer2007} for a related selection bias. We find that this bias results in variability-selected $M_{\rm{BH}}$ values that are on average larger by $\sim 1$ dex than expected from the \citet{Reines2015} relation for $M_{\star} < 10^{9}\ M_{\odot}$ host galaxies. This bias is only slightly reduced with more photo-metrically sensitive light curves. We therefore expect variability-selected IMBH candidates in dwarf galaxies to be strongly affected by this bias. This demonstrates the importance of obtaining additional $M_{\rm{BH}}$ estimates for variability-selected AGNs, such as from the variability timescale \citep{Burke2021c} or broad emission line signatures \citep{Shen2013}, rather than using the stellar mass alone as a proxy.

\subsection{Extension beyond the local Universe}

We have shown that the number of detectable IMBHs falls off quickly with redshift (Figure~\ref{fig:massredshift}) faster than the gain in volume. However, extensions of our model beyond the local Universe are straightforward if one is interested in AGNs with somewhat larger BH masses, $M_{\rm{BH}} \sim 10^5-10^6\ M_{\odot}$, that are detectable at intermediate redshifts (e.g., \citealt{Guo2020,Burke2021b}). To extend the treatment to higher redshifts, one could adopt the same GSMF form of Equation~\ref{eq:GSMF}, but adjust the parameters based on the redshift range using observational constraints on the GSMF evolution (e.g., \citealt{Marchesini2009,Adams2021}). A model for the commensurate host-galaxy $K$-correction (e.g., \citealt{Chilingarian2010}) to the mass-to-light ratios would need to be considered. At intermediate redshifts, the host galaxy-BH mass relation may have a different normalization and slope that better describes the AGN population (e.g., \citealt{Caplar2018,Ding2020}). Obviously, the GSMF in the dwarf galaxy regime becomes less well-constrained with increasing redshift. In addition, whether and how the ERDF of the obscured AGN fraction changes with redshift is uncertain at present. Finally, there are other factors (e.g., dwarf galaxy-galaxy mergers) that complicate using occupation fraction as a direct tracer of seeding scenarios at high redshift \citep{Volonteri2010,Ricarte2018,Mezcua2019,Buchner2019}. Investigations of IMBH evolution in dwarf galaxies using cosmological simulations that incorporate the relevant physics on these scales may help illuminate the properties of the evolving IMBH population \citep{Sharma2022,Haidar2022}.

\subsection{Caveats \& Future work}

Our methodology can be extended and applied to other wavelengths, such as sensitive X-ray observations of dwarf galaxies with eROSITA \citep{Predehl2021,Latimer2021} or time-domain UV imaging surveys \citep{Sagiv2014,Kulkarni2021}. Better constraints on the shape and normalization of the ERDF in the IMBH regime would help us compute our forecasts for the total number of detectable variable dwarf AGNs. Ultimately, a variety of multi-wavelength probes are desired to derive robust constraints on the occupation fraction. 

Though counter-intuitive, it has been amply demonstrated by many previous workers including \cite{Ricarte2018} that local observations of the occupation fraction of black holes in low mass dwarf galaxies could serve to discriminate between high redshift initial seeding models. Despite the fact that post-seeding black hole growth occurs via  accretion and mergers over cosmic time, the memory of these initial seeding conditions may yet survive, in particular, for these low mass galaxies that preferentially host IMBHs. And while current observations cannot conclusively discriminate between alternative initial seeding models as yet, the prospects for doing so are promising as we describe below.

Our modeling indicates that the ``light'' seeding scenario is slightly more consistent with current observational constraints from dwarf AGN variability, however, the current observational constraints in the dwarf galaxy regime (Figure~\ref{fig:varfrac}) are not particularly strong. The discriminating power of optical variability to distinguish between seeding scenarios lies in the capability to accurately measure the variable detected fraction in $M_{\star} \lesssim 10^8 M_{\odot}$ galaxies. Our model predictions for the occupation fractions in scenario (i) and (ii) can be differentiated at the $2-3 \sigma$ level in the detectable variable fraction at $M_{\star} \lesssim 10^8 M_{\odot}$ (see Figure~\ref{fig:varfrac}). Therefore, we are unable to strongly rule-out either seeding scenario (or a mixture of several) at this time except for ones that predict occupation fractions of zero in dwarf galaxies. The large uncertainties here are dominated by uncertainties in the GSMF, optical variability properties, and scatter in the host-mass scaling relation. We expect constraints on some of these quantities to improve dramatically in the near future.

We encourage theoretical developments investigating how the occupation fraction or number density of wandering BHs could correlate with the host galaxy stellar mass would allow us to make predictions for the variable fractions of that population (Figure~\ref{fig:varfrac}). We have shown that a variable wandering IMBH population could be probed with LSST Rubin. This could yield crucial insights to seeding scenarios and the dynamics of IMBHs within galaxies.

We have made some assumptions in our model using the average properties of the galaxy population to predict variability amplitudes. For example, the predicted observed variability amplitudes in our model depend on our population-level model of host galaxy color index and the level of contamination in the light curve aperture. In order to eliminate these assumptions, one could directly use catalog properties, e.g. measured host galaxy luminosities within light curve apertures, from the parent sample of the observations as long as one is cautious about the relevant selection biases in the parent sample properties. Additionally, we caution that the \citet{MacLeod2010,Suberlak2021,Burke2021} parameters are likely to be affected by selection biases, and whether these relations hold in the ADAF/RIAF regime is also somewhat uncertain. 

Nevertheless, we have demonstrated the expected capabilities and prospects of the LSST Rubin wide-fast-deep survey for IMBH identification via optical variability. With robust observational constraints, the problem could be turned around to become an inference problem to constrain the multiple free parameters in our model with priors derived from observational constraints \citep{Caplar2015,Weigel2017}. Improved constraints on the optical variability properties in the IMBH regime will further reduce the uncertainties. Additionally, a wide-field, deep, flux limited catalog of stellar masses of low-redshift galaxies is urgently needed in the Southern Hemisphere to obtain enough statistical power to distinguish between seeding mechanisms with LSST Rubin. Finally, the variability timescale recovery analysis of \S\ref{sec:varrecovery} could be extended or a metric developed to aid in optimization of survey cadences for IMBH discovery.

\subsection{A note on the optical variability amplitude}

The arguments in \S\ref{sec:var} could pose a quantifiable, unified interpretation of the nuclear optical variability amplitude of galaxies and AGNs where the intrinsic variability amplitude is set by the accretion rate and BH mass, but the resulting observed variability amplitude is diluted by the host galaxy emission. This approach provides  quantitative phenomenological predictions for IMBH optical variability, which is argued to show fast and small amplitude variability (e.g., \citealt{Martinez-Palomera2020}).

\section{Conclusions} \label{sec:conclusions}

We have investigated prospects for IMBH discovery using optical variability with LSST Rubin by building a forward Monte Carlo model beginning from the galaxy stellar mass function. After assuming several possibilities for the BH occupation fraction, and incorporating observed galaxy-BH scaling relations, we demonstrate our model's capability to reproduce existing observations. Below, we summarize our main conclusions:
\begin{enumerate}
    \item We confirm the discriminating power of optical variability to distinguish between BH occupation fractions by accurately measuring the variable detected fraction in the $M_{\star} \sim 10^6\ - 10^8 M_{\odot}$ regime.
    
    \item Current observational constraints are however, insufficient to constrain early seeding scenarios given their limited statistical power and the theoretical uncertainties in this regime. However, they are inconsistent with an IMBH occupation fraction of zero near $M_{\star} \sim 10^8 M_{\odot}$.
    
    \item We demonstrate the resulting BH masses may be biased toward larger $M_{\rm{BH}}$ on average at fixed $M_{\star}$ from an Eddington-type bias, depending on the photometric precision of the survey.
    
    \item Given these findings, we forecast detection of up to $\sim 10^2$ IMBHs with LSST Rubin using optical variability assuming an optimistic ``light'' seeding scenario and perhaps more if there exists a population of wandering IMBHs with an Eddington ratio distribution similar to that of SMBHs in red galaxies. 
    
    \item A targeted $\sim$ hourly cadence program over a few nights can provide constraints on the BH masses of IMBHs given their expected rapid variability timescales.
\end{enumerate}


%

\section*{Acknowledgements}

We thank the anonymous referee for a careful review and assessment that has improved our paper. We thank Chris Done and Rodrigo Nemmen for helpful discussions. We thank Konstantin Malanchev and Qifeng Cheng for referring us to an improved algorithm for generating DRW time series. CJB acknowledges support from the Illinois Graduate Survey Science Fellowship. YS acknowledges support from NSF grant AST-2009947. XL and ZFW acknowledge support from the University of Illinois Campus Research Board and NSF grants AST-2108162 and AST-2206499. PN gratefully acknowledges support at the Black Hole Initiative (BHI) at Harvard as an external PI with grants from the Gordon and Betty Moore Foundation and the John Templeton Foundation.  This research was supported in part by the National Science Foundation under Grant No. NSF PHY-1748958. This research made use of Astropy,\footnote{\url{http://www.astropy.org}} a community-developed core Python package for Astronomy \citep{astropy2018}.

\section*{Data Availability}

The data used in this work is available following the references and URLs in the text. Our pre-computed SED templates are available at \url{https://doi.org/10.5281/zenodo.6812008}.



\bibliographystyle{mnras}
\bibliography{seed} 




\appendix

\section{Comparison of mock sample to SDSS}\label{sec:nsacomp}

\begin{figure*}
\centering
\includegraphics[width=0.98\textwidth]{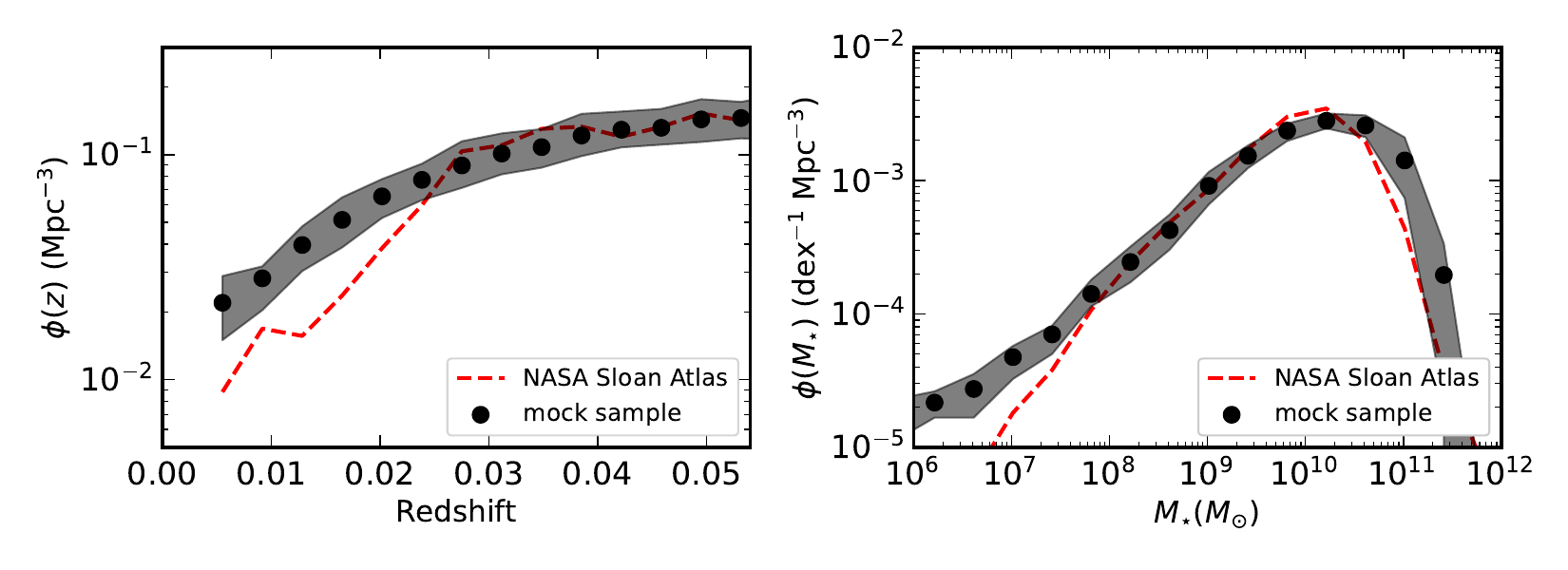}
\caption{Distributions of galaxy redshifts (\emph{left}) and stellar masses (\emph{right}) for a mock sample with a limiting magnitude of $r\approx17.8$ (black circle symbols) and the NASA Sloan Atlas catalog of $z<0.055$ SDSS galaxies (dashed red line). Our mock sample reproduces the observed distributions from the NASA Sloan Atlas reasonably well. \label{fig:redshift}}
\end{figure*}

As an additional check to ensure that our mock sample of galaxies have reasonable properties, we plot the redshift and stellar mass number densities of our mock sample and the real data from the NASA Sloan Atlas catalog of $z<0.055$ SDSS galaxies, based on the SDSS data release 8 \citep{Blanton2011} with $\Omega\approx9380$ deg$^2$. To ensure the comparison is consistent, we apply magnitude limits of $r\approx17.8$ mag to each sample using the spectroscopic targeting limit of SDSS after applying the rough filter conversions from Lupton (2005)\footnote{\url{http://www.sdss3.org/dr8/algorithms/sdssUBVRITransform.php\#Lupton2005}}. We also assume a measurement uncertainty of 0.3 dex in stellar mass for our mock sample \citep{Reines2015}. The result is shown in Figure~\ref{fig:redshift}. As expected, the mock sample and real sample have qualitatively similar redshift and mass distributions.

\section{Host Galaxy Dilution Parameters}\label{sec:host}

\begin{figure*}
\raggedleft
\includegraphics[width=1\textwidth]{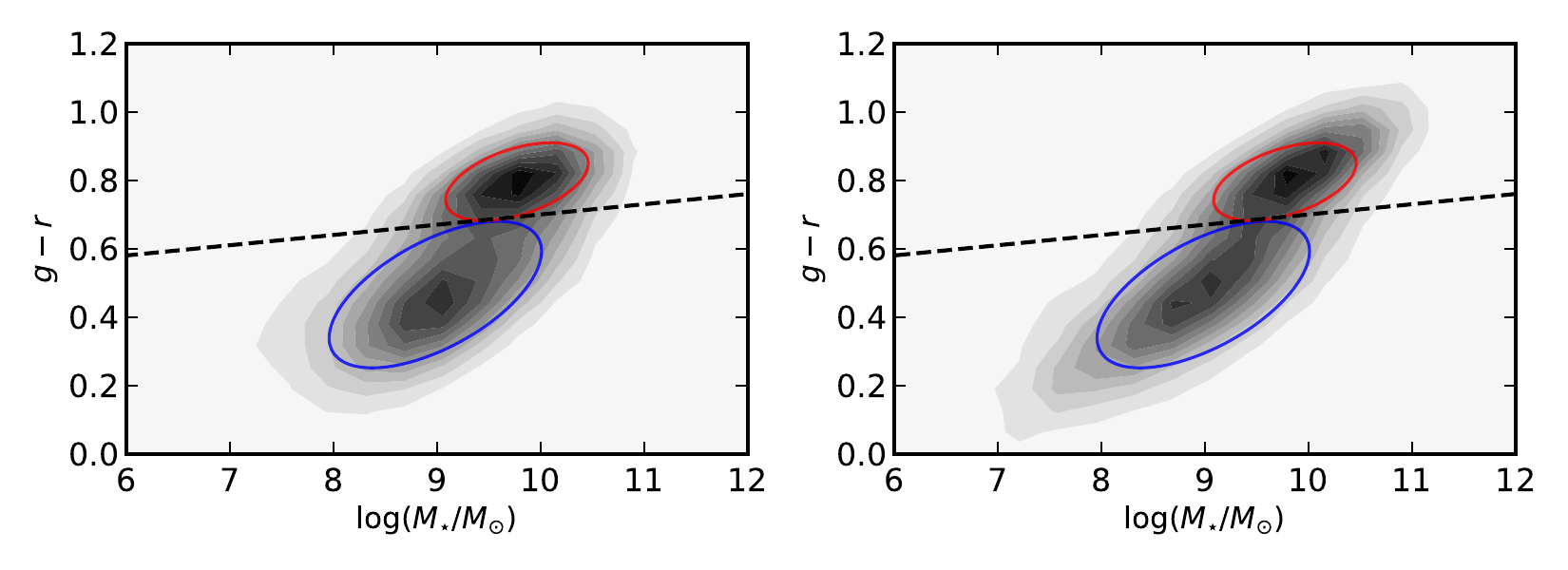}
\caption{Host galaxy $g-r$ color index versus stellar mass from the NASA Sloan Atlas catalog of $z<0.055$ SDSS galaxies (\emph{left panel}) and for a mock sample with a limiting magnitude of $r\approx17.8$ (\emph{right panel}). Darker contours represent denser regions of the distributions. The exact shapes of the contours depend on the limiting magnitude. The red and blue ellipses are the $1\sigma$ contours of the two Gaussian distributions for the red and blue galaxy populations fit to the SDSS data, from which we randomly draw the galaxy colors in our Monte Carlo model. They are shown in both panels to facilitate comparison. The dashed black line shows the color-magnitude diagram slope of $-0.03$ used to divide the red and blue galaxy populations \citep{Bell2003}. \label{fig:hostcolor}}
\end{figure*}

\begin{figure*}
\raggedleft
\includegraphics[width=1\textwidth]{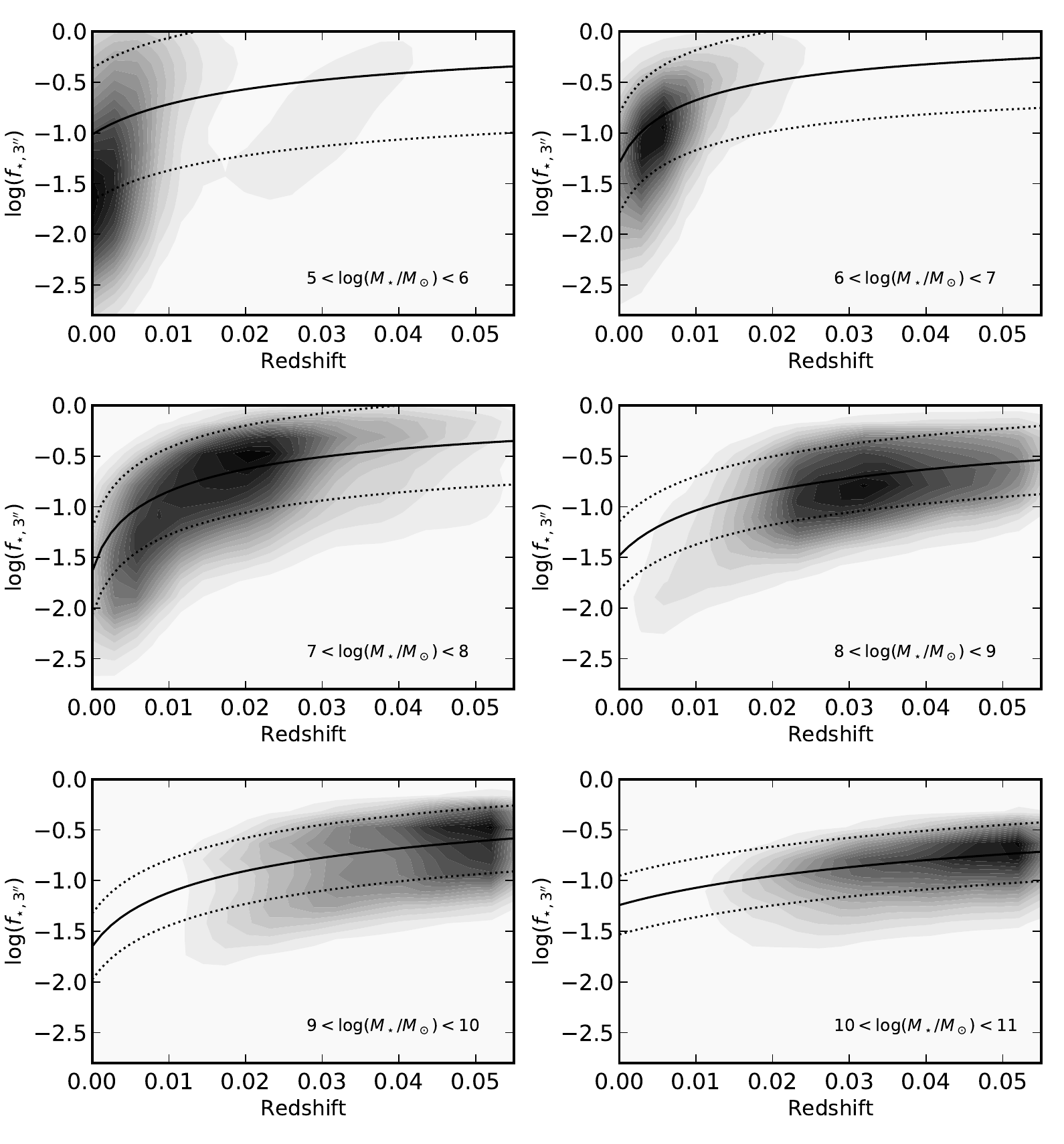}
\caption{Host galaxy aperture contamination factor in a $3^{\prime\prime}$-diameter aperture $f_{\star, 3^{\prime\prime}}$ versus redshift from the NASA Sloan Atlas catalog of $z<0.055$ SDSS galaxies in bins of stellar mass (marked on each panel). Darker contours represent denser regions of the distributions. Our adopted best-fitting model and rms scatter are shown as solid and dotted black lines, respectively. \label{fig:hostf}}
\end{figure*}

We compute the host galaxy Petrosian $g-r$ color index from the version 0.1.2 of the NASA Sloan Atlas catalog of $z<0.055$ SDSS galaxies, based on the SDSS data release 8 \citep{Blanton2011}. In order to sample realistic colors for our galaxies in our model, we incorporate the bi-modality of the galaxy color population (e.g., \citealt{Bell2003,Baldry2004}). Because we are interested in observed colors at low redshifts, we use the observed Petrosian magnitudes without K-corrections or dust-corrections using the \texttt{PETROFLUX} key. We model the color-mass diagram ($g-r$ versus stellar mass) as a mixture of elliptical Gaussians, following a similar approach to \citet{Taylor2015}. We used the \texttt{BayesianGaussianMixture} module implementation in \textsc{scikit-learn} python package \citep{scikit-learn}. The $g-r$ color indices are then sampled using these probability distribution functions at a given stellar mass for the red and blue galaxy populations separately using the respective GSMF (\S\ref{sec:GSMF}). The two Gaussian components representing the red and blue galaxy populations are shown on the color-mass diagram in Figure~\ref{fig:hostcolor}. A typical $g-r$ color index value for a radiatively-efficient (blue) dwarf galaxy is $g-r\approx0.5$.

We also compute the aperture contamination factor (covering factor) accounting for the level of host galaxy light dilution in a $3^{\prime\prime}$ aperture, denoted $f_{\star, 3^{\prime\prime}}$. We obtain these values by dividing the flux within a circular $3^{\prime\prime}$ aperture by the total Petrosian flux as $\texttt{FIBERFLUX}/\texttt{PETROFLUX}$ in the $g$ band. There are two effects to consider. First, the aperture contamination increases with redshift as the typical galaxy angular size decreases. Second, the aperture contamination increases as galaxy stellar mass decreases given the galaxy the size-mass relation (e.g., \citealt{vanderWel2014}). Therefore, we split the SDSS galaxies into bins of stellar mass and evaluate the $f_{\star, 3^{\prime\prime}} - z$ relations in each bin. We fit an empirical polynomial function of the form $f(x) = 1 - 1/(x^2 + bx + c)$ where $x\equiv z-a$, which assures that $f(x) \xrightarrow{} 1$ as $z \xrightarrow{} \infty$, to the distribution of $f_{\star, 3^{\prime\prime}}$ versus redshift $z$. The scatter in these relations is probably a function of the varying surface brightness profiles in the galaxy population. For example, galaxies with bright bulges will have larger $f_{\star, 3^{\prime\prime}}$ values. We adopt this simple best-fitting models and rms scatter in our Monte Carlo framework. The results for both are shown in Figure~\ref{fig:hostf}. A typical host galaxy dilution parameter value for dwarf galaxies near the median redshift of $z\approx0.03$ is $f_{\star, 3^{\prime\prime}}\approx0.2$.

\section{Black hole mass detection limits}\label{sec:appdetlim}

\begin{figure}
\centering
\includegraphics[width=0.5\textwidth]{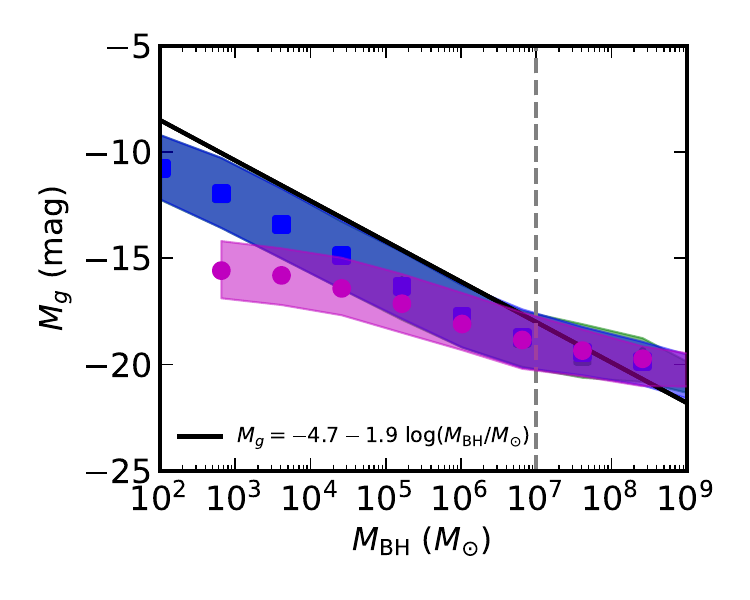}
\caption{Combined host galaxy and AGN absolute magnitude - BH mass relation for our LSST Rubin (lower panel) -like models. The solid black line is a linear approximation of the 68th percentile to approximate the faint end of the population for use in computing the detection limits. The gray dashed lines represents the rough transition mass of $M_{\rm{BH}} \sim 10^7\ M_{\odot}$, below which the luminosity is significantly affected by the host galaxy light. \label{fig:Mmass}}
\end{figure}

\begin{figure}
\centering
\includegraphics[width=0.5\textwidth]{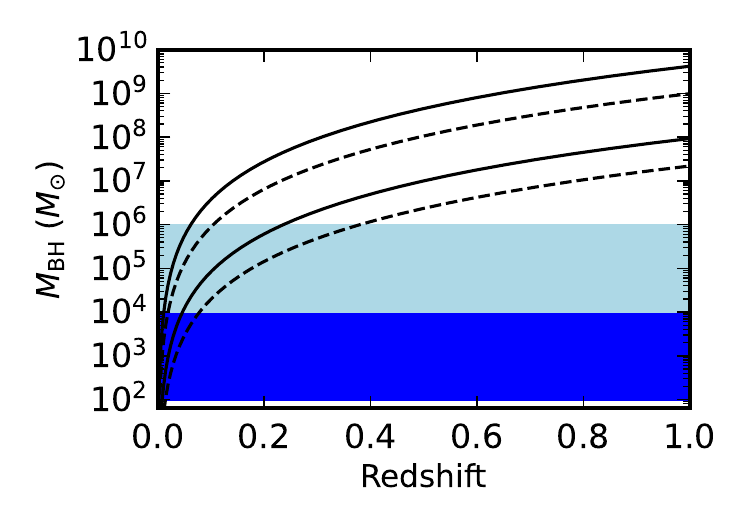}
\caption{Theoretical BH mass detection limit for our PTF (upper pair of black curves) and LSST Rubin (lower pair of black curves) -like models assuming an rms variability amplitude of 0.1 (solid lines) and 0.3 mag (dashed lines). The light blue shaded area represents the range of ``massive BHs'' where the BHMFs begin to differ. The dark blue shaded area represents the IMBH regime.   \label{fig:detlim}}
\end{figure}

A light curve with detectable variability must have an rms variability amplitude greater than the photometric precision limit of the survey. For a sufficiently long light curve ($t_{\rm{baseline}} \gtrsim \tau$), the rms should approximate the asymptotic variability amplitude (host-diluted) $\rm{SF}_\infty^\prime$. The detection limit is given by equating this with the photometric precision of the survey (Equations~\ref{eq:ppm1} and \ref{eq:ppm2}). Assuming the systematic component of the photometric precision is small (generally true at faint magnitudes), and ignoring the (small) first order term of Equation~\ref{eq:ppm2} (i.e., $0.04 - \gamma \approx 0$), we have:
\begin{equation}
    {\rm{SF}}_\infty^\prime \approx \gamma^{1/2}\ x.
\end{equation}
Substituting $x \equiv 10^{0.4\ (m-m_5)}$ above, we have:
\begin{equation}
    {\rm{SF}}_\infty^\prime \approx \gamma^{1/2}\ 10^{0.4\ (m-m_5)}.
\end{equation}
Taking $m = M + 5\ \log d_{\rm{pc}} - 5 + K(z)$ where $K(z)$ is the K-correction, and re-arranging:
\begin{equation}
    2.5\ \log({\rm{SF}}_\infty^\prime\ \gamma^{-1/2}) \approx M + 5\ \log d_{\rm{pc}} - 5 + K(z) - m_5.
\end{equation}
Here, the absolute magnitude $M$ refers to the total magnitude of the host galaxy and the AGN. Noting that the BH mass and galaxy luminosity are correlated with some scatter determined by the scatter in both the mass-to-light ratios and host galaxy - BH mass relation, we assume the absolute magnitude can be described as by linear function of the form $M = a + b\ \log(M_{\rm{BH}}/M_\odot)$ for small BH masses where the host galaxy light dominates, as shown in Figure~\ref{fig:Mmass}. In the $g$-band, we find $a=-4.7$ and slope $b=-1.9$. In $R$, an intercept of $a=-6.0$ is more appropriate. Substituting above and solving for $\log(M_{\rm{BH}}/M_\odot)$, we have:
\begin{multline}
    \log\left(\frac{M_{\rm{BH}}}{M_\odot}\right) \approx \frac{1}{b} \biggl[ 2.5\ \log({\rm{SF}}_\infty^\prime\ \gamma^{-1/2}) - a - 5\ \log d_{\rm{pc}} + 5 \\ - K(z) + m_5 \biggr].
\end{multline}
We show the BH mass detection limits for our PTF and LSST Rubin -like models in Figure~\ref{fig:detlim}, taking ${\rm{SF}}_\infty^\prime = 0.1$ and ${\rm{SF}}_\infty^\prime = 0.3$ mag (Figure~\ref{fig:SFinf}). There is likely to be a complex mass dependence on the intrinsic variability amplitude, depending on the intrinsic BH mass dependence on the variability amplitude and the host galaxy luminosity (see \S\ref{sec:var}), so we adopt these two scenarios as a simplification. The K-correction is assumed to be zero, because it is usually small for blue, star-forming dwarf galaxies \citep{Chilingarian2010}. The corresponding theoretical stellar mass detection limit can then be computed using the BH - host galaxy stellar mass relation.

\section{Variable fraction for the red and blue host galaxy populations}\label{sec:appredblue}

\begin{figure*}
\centering
\includegraphics[width=0.98\textwidth]{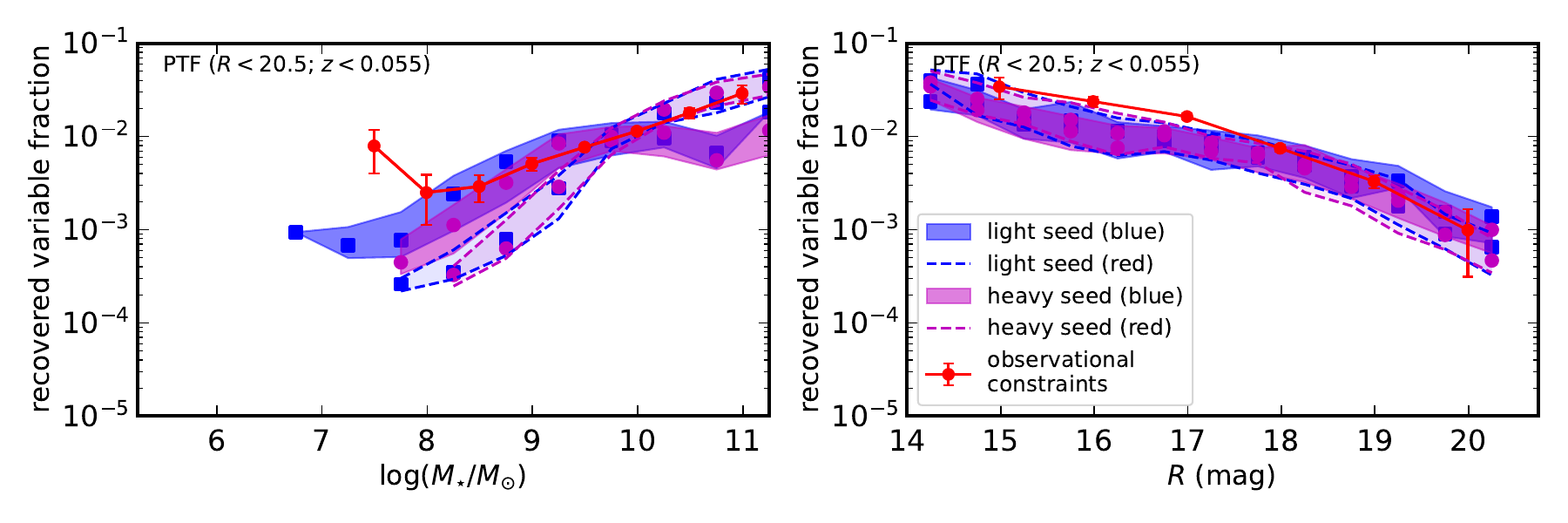}
\includegraphics[width=0.98\textwidth]{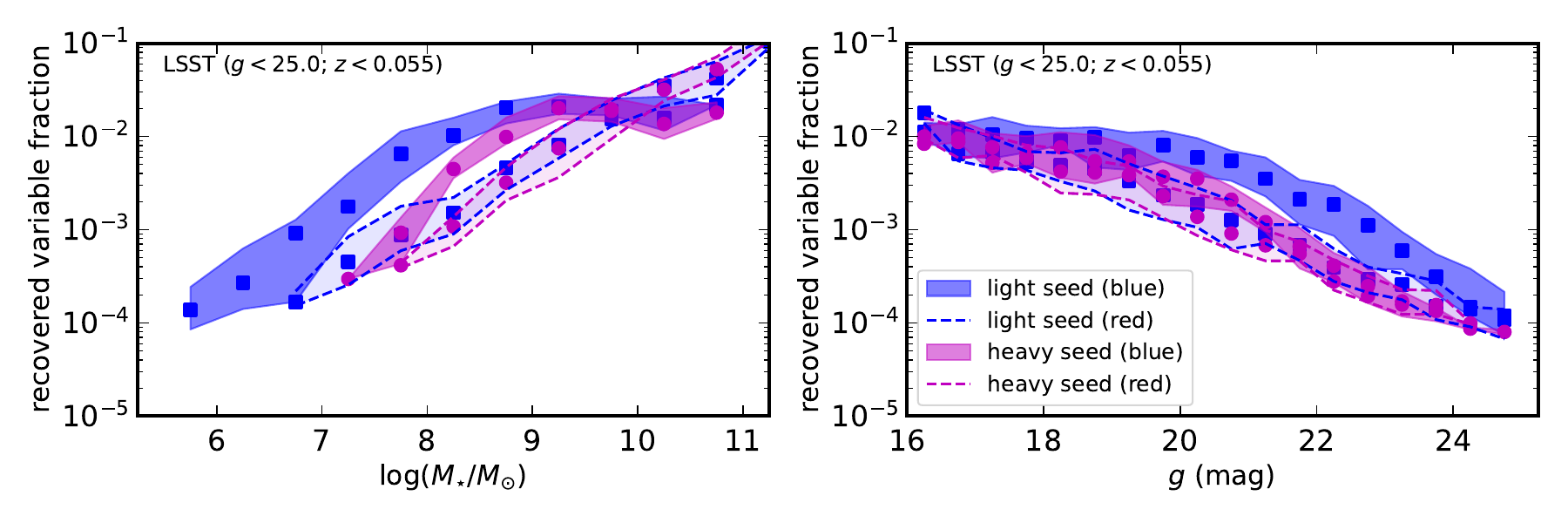}
\caption{Same as the lower panel of Figure~\ref{fig:varfrac} but with the variable fractions computed separately for the blue (shaded bands) and red (lightly shaded bands with dashed lines) host galaxy populations as a fraction of the total host galaxy population. \label{fig:appvarfractype}}
\end{figure*}

Following \S\ref{sec:syserr}, we show versions of Figure~\ref{fig:varfrac} for the blue and red host galaxy populations with different ERDFs \citep{Weigel2017} with a 0.3 dex uncertainty on the stellar mass measurements. The results are shown in Figure~\ref{fig:appvarfractype} for our PTF and LSST Rubin -like models. Red host galaxies make up a larger fraction of the total variable AGNs with $M_{\star} \gtrsim 10^{9.5}\ M_{\odot}$ while blue host galaxies dominate the variable dwarf AGNs with $M_{\star} \lesssim 10^{9.5}\ M_{\odot}$.

\section{Variable fraction for varying uncertainty on stellar mass measurements}\label{sec:appsyserr}

\begin{figure*}
\centering
\includegraphics[width=0.98\textwidth]{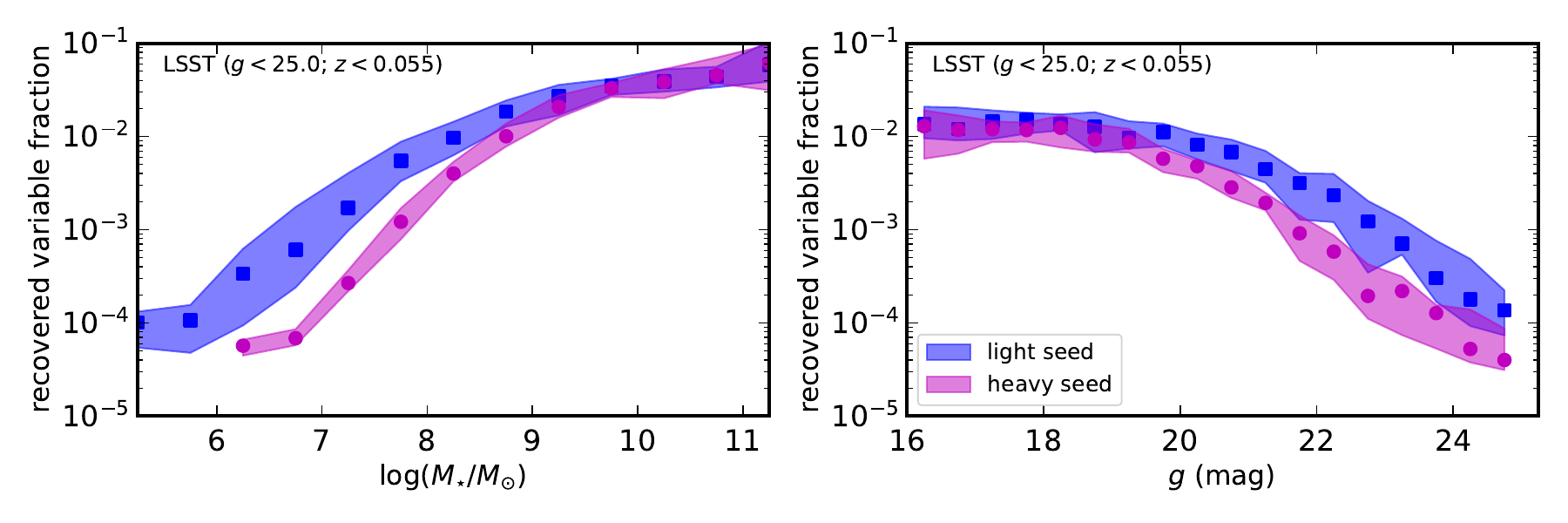}
\includegraphics[width=0.98\textwidth]{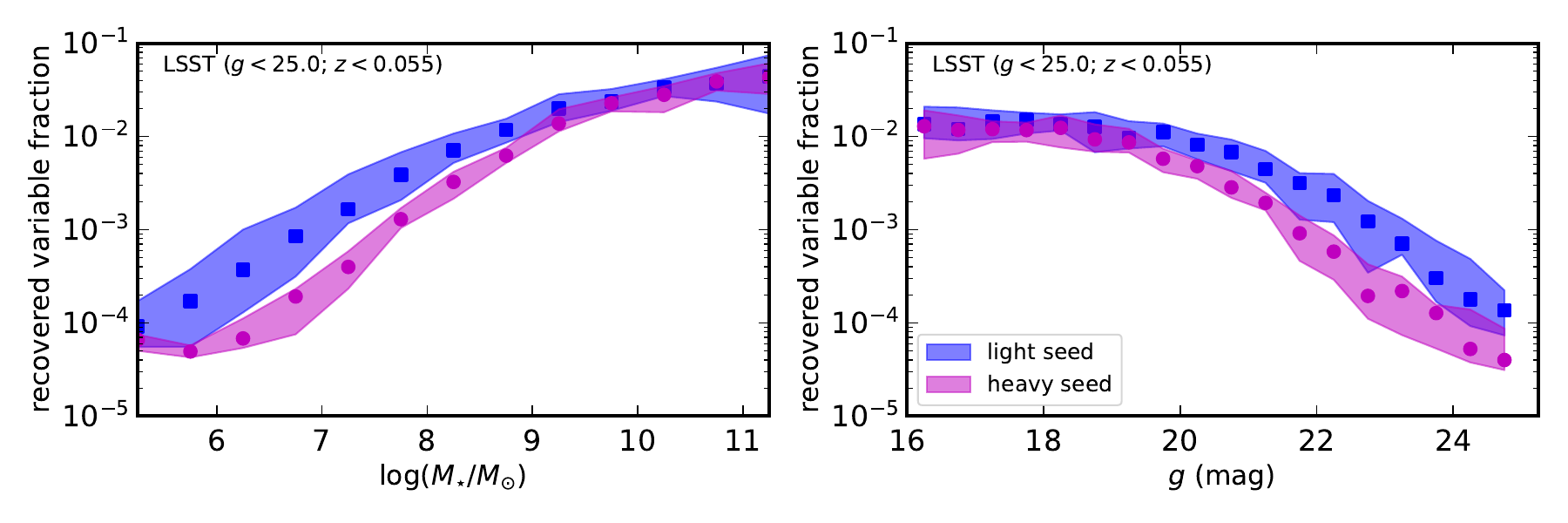}
\caption{Same as the lower panel of Figure~\ref{fig:varfrac} but with 0.6 dex (\emph{upper panel}) and 0.9 dex (\emph{lower panel}) uncertainties on the stellar mass measurements. \label{fig:appvarfrac}}
\end{figure*}

Following \S\ref{sec:syserr}, we show versions of Figure~\ref{fig:varfrac}, which assumed a 0.3 dex uncertainty on the stellar mass measurements, with 0.6 dex and 0.9 dex uncertainties on the stellar mass measurements for comparison. The results are shown in Figure~\ref{fig:appvarfrac} for our LSST Rubin -like models. The stellar mass uncertainties have no effect on the variable fraction as a function of magnitude.

\bsp	
\label{lastpage}
\end{document}